\documentclass[%
 reprint,
 showpacs,
 amsmath,amssymb,
 aps,
]{revtex4-1}

\usepackage{graphicx}
\usepackage{dcolumn}
\usepackage{bm}

\newcommand{\sakujo}[1]{}

\newcommand{\HALL}{\alpha}
\newcommand{\HEL}{\sigma}
\newcommand{\POL}{s}

\newcommand{\U}{\underline}

\newcommand{\Elsasser}{Els\"asser}

\newcommand{\LieBrace}[3]{{#1\{}#2,#3{#1\}}}
\newcommand{\LieBracket}[3]{{#1[}#2,#3{#1]}}

\newcommand{\Braket}[3]{{#1<}#2{#1|}#3{#1>}}
\newcommand{\BraKet}{\Braket}
\newcommand{\ParKet}[3]{{#1(}#2{#1|}#3{#1)}}

\newcommand{\GEV}[2]{\vec{\bm{#1}}#2}
\newcommand{\GEC}[2]{\widehat{#1}#2}
\newcommand{\GEM}[2]{\tilde{#1}#2}

\newcommand{\CHW}[2]{{\bm{#1}}#2}
\newcommand{\CHC}[2]{\widehat{#1}#2}

\newcommand{\tripleHMHD}[4]{{#1(}\!{#1(}#2{#1|}\!{#1|}#3{#1|}\!{#1|}#4{#1)}\!{#1)}}

\newcommand{\CC}[1]{\overline{#1}}

\newcommand{\dd}[2]{\frac{\partial #2}{\partial #1}}
\newcommand{\sdd}[2]{\frac{{\rm{d}} #2}{{\rm{d}} #1}}

\newcommand{\CalcNote}[1]{}%
\renewcommand{\CalcNote}[1]{*>#1<*}%
\renewcommand{\sakujo}[1]{/*#1*/}%

\newcommand{\curl}{\nabla\times}

\newcommand{\Vi}[1]{\bm{U}_{#1}}
\newcommand{\Ve}[1]{\bm{J}_{#1}}
\renewcommand{\Vi}[1]{\bm{V}_{i#1}}
\renewcommand{\Ve}[1]{\bm{V}_{e#1}}
\newcommand{\Vx}[1]{\bm{V}_{x#1}}
\newcommand{\Mi}[1]{\bm{M}_{i#1}}
\newcommand{\Me}[1]{\bm{M}_{e#1}}
\newcommand{\Mx}[1]{\bm{M}_{x#1}}

\newcommand{\EIGK}{\Lambda}

\begin{document}

\preprint{APS/123-QED}

\title{%
Helicity-based, particle-relabeling operator and normal mode expansion
of the dissipationless incompressible Hall magnetohydrodynamics
}

\author{Keisuke Araki}
 \email{araki@are.ous.ac.jp}
\affiliation{%
 Faculty of Engineering, Okayama University of Science,
 1-1 Ridai-cho, Kita-ku, Okayama 700-0005 JAPAN
}%
%
%


\date{\today}

\begin{abstract}
The dynamics of 
an incompressible, dissipationless Hall magnetohydrodynamic medium
are investigated from Lagrangian mechanical viewpoint.
The hybrid and magnetic helicities are shown to emerge, respectively,
from the application of the particle relabeling symmetry 
for ion and electron flows 
to Noether's first theorem,
while
the constant of motion associated with the theorem is generally given by
their arbitrary linear combination.
Furthermore,
integral path variation associated with the invariant action 
is expressed by the operation of an integro-differential operator 
on the reference path.
The eigenfunctions of this operator are double Beltrami flows,
i.e. force-free stationary solutions to the equation of motion
and
provide a family of orthogonal function bases
that yields the spectral representation of the equation of motion
with a remarkably simple form.
Among the double Beltrami flows,
considering 
the influence of a uniform background magnetic field
and 
the Hall term effect vanishing limit%
,
the generalized {\Elsasser} variables 
are found to be the most suitable for avoiding 
problems with singularities in the standard magnetohydrodynamic limit.
%
%
\end{abstract}

\maketitle


\section{Introduction}

In the present study 
we investigate
dynamical system features of 
a dissipationless incompressible Hall magnetohydrodynamic (HMHD) medium
and propose
the notion of helicity-based, particle-relabeling operator,
which is located at the junction of two seemingly separated topics:
particle relabeling symmetry and force-free, stationary state solution.
Consideration of
the invariant action associated with the particle-relabeling symmetry
naturally leads to the operator,
and
its eigenvalue problem and associated normal-mode expansion of
basic formulas and equations are examined as its application.

\vspace{1em}

The HMHD is well-known 
as a simple, one-fluid plasma model that
contains two-fluid effects
and
that has been intensively investigated
both numerically and mathematically.
The basic idea of the HMHD approximation is formulated 
by replacing the magnetohydrodynamic (MHD) approximation
\footnote{%
Despite that the term ``magnetohydrodynamics'' fundamentally means 
continuous fluid approximation models of 
the collective motions of ions and electrons
\cite{miyamoto2000fundamentals},
in many literatures the term is used to indicate 
the reduced, one-fluid plasma model wherein
the ion and electron momenta are averaged,
the (averaged) ion charge number, $Z$, is often set to one,
and
the Lorentz force is formulated using these averaged quantities.
In the present study we use the term ``MHD'' or ``standard MHD'' 
in this meaning.
},
by which a vanishing Lorentz force is assumed for the entire plasma
($\bm{E}+\bm{V}\times\bm{B}=\bm{0}$),
with an assumption that
the Lorentz force only vanishes for the electron component of the plasma
($\bm{E}+\bm{V}_{e}\times\bm{B}=\bm{0}$),
where $\bm{V}$ and $\bm{V}_{e}$ are 
the averaged plasma velocity and 
its electron component, respectively
\cite{Lighthill1960}.
The formulation is completed by
approximating the entire plasma velocity by its ion component
($\bm{V}\approx\bm{V}_{i}$),
and then
evaluating the current density by
$\Vi{}-\Ve{}=\bm{J}/en_{e}$
with
an approximated Ampere's law,
$\nabla\times\bm{B}=\mu_{0}\bm{J}$.
Thus,
the evolution equations for 
a dissipationless, incompressible HMHD plasma
are given by 
the incompressibility condition,
solenoidal condition,
momentum equation,
and
induction equation
as follows:
\begin{eqnarray}
&&
  \nabla\cdot\bm{u}
  =
  \nabla\cdot\bm{b}
  =
  0
,
\nonumber
\\&&
  \partial_{t}{\bm{u}}
  =
  \bm{u}\times(\nabla\times\bm{u})
  +
  \bm{j}\times\bm{b}
  -
  \nabla P
,
\label{eq of mot (u,b)}
\\&&
  \partial_{t}{\bm{b}}
  =
  \curl\big((\bm{u}-\HALL\bm{j})\times\bm{b}\big)
,
\nonumber
\end{eqnarray}
where $\bm{u}$, $\bm{b}$, $\bm{j}$, $P$, and $\HALL$ 
are the appropriately nondimensionalized variables corresponding to
ion velocity, magnetic field, current density ($\bm{j}=\curl\bm{b}$), 
generalized pressure, and Hall term strength parameter, respectively.
It is easy to see that, in the limit $\HALL\to0$, 
the system reduces to the standard MHD system.

The system (\ref{eq of mot (u,b)}) 
is known to have three constants of motion,
i.e., the total energy, $E$, the magnetic helicity, $H_M$,
and the hybrid helicity, $H_H$, which are given by
\begin{eqnarray}
&&
  E=\frac12\int(|\bm{u}|^2+|\bm{b}|^2){\rm{d}}^3\vec{x}
,
\label{total energy}
\\&&
  H_M=\frac12\int\bm{a}\cdot\bm{b}{\rm{d}}^3\vec{x}
,
\label{magnetic helicity}
\\&&
  H_H=\frac12\int(\HALL\bm{u}+\bm{a})\cdot
  (\HALL\curl\bm{u}+\bm{b}){\rm{d}}^3\vec{x}
,
\label{hybrid helicity}
\end{eqnarray}
respectively \cite{Turner1986},
where $\bm{a}$ is the vector potential of $\bm{b}$ ($\bm{b}=\curl\bm{a}$).
Obviously, in the MHD limit, $\HALL\to0$,
the hybrid helicity degenerates into the magnetic helicity. 
However, it was very interesting that 
the spectral representation of (\ref{eq of mot (u,b)})
by the generalized {\Elsasser} variables was naturally proved to yield 
\textit{four} constants of motion
due to the skew-symmetry of the quadratic terms coefficients
\cite{Araki2015}.
The fourth constant was the modified cross helicity, 
given by $H_C:=H_H-H_M$,
and it converges to the cross helicity in the MHD limit.
Despite that 
the helicity conservation 
has been known to emerge from 
the particle relabeling symmetry 
for the MHD case \cite{PadhyeMorrison1996},
its HMHD counterpart still remains unresolved.

We focus here on the Lagrangian mechanical aspects of the HMHD system
together with a differential topological framework.

Since
Holm established the Hamiltonian mechanical description of the HMHD system
\cite{Holm1987},
analytical mechanical approaches to HMHD physics
have been mainly carried out
within the Hamiltonian mechanics framework
\cite{SahraouiETAL2003,HirotaETAL2006}.
Recently,
a Lagrangian mechanical approach was employed by
Keramidas Charidakos et al. \cite{KeramidasCharidakosETAL2014}.
Their Lagrangian was obtained by 
naturally extending 
an $n$-particle system Lagrangian
to a two-fluid plasma model,
and
the HMHD momentum equation and Ohm's law 
were derived.

On the other hand,
our Lagrangian mechanical approach is rather close to 
Arnold's differential-geometrical method.

Ever since Arnold \cite{Arnold1966}
reviewed his studies of dynamical systems on Lie groups 
and related hydrodynamic topics in a unified form, 
many fluid dynamical systems have been recognized 
to exist on appropriate Lie groups 
\cite{ArnoldKhesin1998}.
The key mathematical objects of Arnold's method are twofold.
One is an appropriate Riemannian metric 
that is introduced on the relevant Lie group
as a Lagrangian of the action.
The other are so-called ``Lin's constraints''
that provide the variation of an integration path
\cite{MarsdenRatiu1994}.

In the field of plasma physics,
Arnold's method was found to be applicable to the dynamics of 
a dissipationless, incompressible MHD medium 
if the Lie algebraic structure was appropriately defined 
on the function space of the pair of the velocity and magnetic fields
\cite{ZeitlinKambe1993,Hattori1994}.
This extension to the pair was called ``magnetic extension''
and
is now recognized 
as a special case of the \textit{semidirect product} of 
a Lie group and a certain vector space
(Sect. 10.B of Ref.\cite{ArnoldKhesin1998}).

Since
the induction equation is not ``passive''
due to the Hall term
(i.e.,
the magnetic field can evolve autonomously),
the HMHD system does not obey 
the magnetic extension scheme.
This discrepancy of magnetic extension can be overcome
by replacing the group action on the vector space
with the group homomorphism,
which was based on Vizman's extended formulation
\cite{Vizman2001}.
The configuration space was given by 
a semidirect product of two volume-preserving diffeomorphisms
and
the Lagrangian was given by a Riemannian metric
that physically implied the total plasma energy
\cite{Araki2015}.
In the present study,
we will report another HMHD formulation,
wherein
the ion and electron velocities are taken as basic variables
and 
the configuration space is given by 
a \textit{direct product} of two volume-preserving diffeomorphisms,
and 
we will discuss the conservation of helicities
as
a consequence of the particle relabeling symmetry of each fluid.
For the MHD case,
the particle relabeling symmetry and its relation to helicity conservation
is discussed by Padhye and Morrison \cite{PadhyeMorrison1996}.

Note that,
despite the simple appearance of the basic equations
(\ref{eq of mot (u,b)}),
analytical mechanical approaches to HMHD systems
raise a small parameter problem
when their relation to the standard MHD limit is considered.
For example,
in the Hamiltonian mechanics approach,
one of the natural choices of vector variables 
is the pair of
the total ion momentum density,
$
 \bm{M}=\rho \bm{v} + R^{-1} a \rho \bm{A},
$
and
the magnetic vector potential, $\bm{A}$,
where 
$\rho$, $\bm{v}$ are the density and velocity of ion component
and
$R/a=\HALL$ in our notation \cite{Holm1987}.
In the limit $\HALL\to0$,
these two variables come close to each other, 
$\bm{M}\approx R^{-1} a \rho \bm{A}$,
and 
manipulation of the small difference $\bm{v}=\bm{M}/\rho - R^{-1} a \bm{A}$
is needed to capture the ion flow.
Recently,
Yoshida and Hameiri proposed a method to treat the MHD limit 
by renormalizing the Lagrangian described 
by some appropriate Clebsch variables 
\cite{YoshidaHameiri2013}.
In the present study,
we will seek another way of avoiding the singularity problem
by choosing an appropriate expansion function set.

In the context of the analysis of fully-developed turbulence,
it was recently shown by direct numerical simulation (DNS)
that
the Hall term effect alters 
the formation tendency of coherent structures
\cite{MiuraAraki2014}.
Formation of tubular structures of currents and enstrophy densities 
at small scales are observed for the HMHD case, 
while sheet-like structures 
are often observed for the standard MHD system.
In addition,
it is interesting that although
both
the Lorentz force term of the ion velocity evolution equation
and 
the Hall term of the magnetic field evolution equation
contain the function $\bm{j}\times\bm{b}$,
their contributions 
to the energy transfer of the kinetic and magnetic energies
were found to be quite different \cite{ArakiMiura2013}.
This suggests that,
for the analysis of basic dynamical features,
it is not sufficient to focus upon the features of a magnetic field alone,
but that their coupling with the velocity field must also be considered.
Thus,
an appropriate coupled base function system is required 
for the DNS or some other practical analysis.

In relation to the coupling of the magnetic and ion velocity fields,
there exist two significant functional categories
to describe the equilibrium states, dynamics, and the stability
of the HMHD system:
the \textit{double Beltrami flow} (DBF);
and
the \textit{generalized {\Elsasser} variable} (GEV).

The notion of DBF was introduced by Mahajan and Yoshida 
in order to extend the concept of a Taylor state
to two-fluid plasma models
\cite{MahajanYoshida1998}.
To derive the stable equilibrium state, 
the DBF was applied to 
the variational calculation to minimize a dissipation function
\cite{YoshidaMahajan2002}.
The DBF was also applied 
as ``dynamically accessible variation,''
which conserves the Casimir invariants,
to analyze the nonlinear stability of the equilibrium state
\cite{HirotaETAL2006}.
In the present study,
the Casimir-preserving nature of the DBFs 
will be considered
based on its Lagrangian mechanical counterpart, i.e., 
Noether's first theorem,
and
the eigenfunctions of the DBF-generating operator 
will be shown to provide a remarkably simple 
expression for the evolution equation.

On the other hand, 
the GEV was introduced by Galtier to formulate the HMHD dynamics 
in the wave/weak turbulence closure analysis framework 
\cite{Galtier2006}.
Though GEVs were developed to describe 
the linear waves that are excited
when a uniform background magnetic field exists,
they can also be used as a set of orthogonal base functions
even when the ambient field is absent.
In the previous study, 
it was shown that
the GEV expansion of the HMHD equation 
naturally yields four conservation laws
due to the symmetric properties of the quadratic term coefficients,
and
it was conjectured that 
this observation might reflect some symmetries intrinsic to the system
\cite{Araki2015}.
Recently, 
we also have applied the GEV decomposition to the DNS data
and
confirmed the mirror symmetry breaking at small scales
\cite{ArakiMiura2015}.
In the present study,
we will review the GEV as a specific example of DBF
and 
discuss its advantages over other DBFs
in relation to the MHD limit.

\vspace{1em}

This paper is organized as follows:
the basics of the Lagrangian mechanical formulation are given in section 2;
variational calculations are carried out in section 3,
where
we derive
the equation of motion from Hamilton's principle
and
the conservation of the helicities from Noether's first theorem;
in section 4, 
the derivation process is reformulated 
using the differential topological terminology,
and
the topological foundations of helicity conservation are discussed.
The DBFs are used as the base functions of the HMHD system
and
the topological basic quantities, i.e., 
the Riemannian metric and the structure constant of the Lie group, 
are given in the section 5;
the influence of a uniform background magnetic field
and
the standard MHD limit of HMHD system are discussed
in section 6;
the section 7 is devoted to discussing the implications of our findings.

\section{Formulation}

In a Lagrangian mechanical description of hydrodynamics,
the basic variable for describing the fluid motion 
is known to be given by 
an $n$-tuple of functions that maps the fluid particles
from one time to another;
we call this variable the ``particle trajectory map'' (PTM) hereafter.

In the present study,
we choose as the basic variables,
a pair of PTMs, say $(\vec{X}(t),\vec{Y}(t))$,
which describe 
the positions of the ion and electron fluid particles, respectively;
$\vec{X}(\vec{a},t)$, $\vec{Y}(\vec{a},t)$
express the positions,
which are initially (at $t=0$) located at $\vec{a}\in M$.%
\footnote{%
In this paper,
we place an arrow above the symbol 
to denote the multifunctional character of mathematical quantities.
For example,
a diffeomorphism (a triplet of functions) 
is expressed by $\vec{X}=(X^{1},X^{2},X^{3})$,
and 
a pair of vector fields by $\GEV{V}{}=(\Vi{},\Ve{})$.
Boldface letters are used to denote vector fields on $M$.
}
In differential topological terminology,
we consider here
the dynamical system 
on a \textit{direct} product of two volume-preserving diffeomorphisms,
say $G:=S{\rm{Diff}}(M) \times S{\rm{Diff}}(M)$, hereafter.
Note that,
the choice of basic variables is not unique for the HMHD system;
in our previous study,
we used the pair of PTMs of 
the ion velocity and the current density 
to constitute a \textit{semidirect} product of diffeomorphisms
\cite{Araki2015}.

The PTMs are related to 
the ion and electron velocity fields 
in the Eulerian specification,
say
$
  \bm{V}_{i}(t)={V}_{i}^{k}(t)\dd{x^{k}}{}
,
$
$
  \bm{V}_{e}(t)={V}_{e}^{k}(t)\dd{x^{k}}{}
  \in
  \mathfrak{X}_{\Sigma}(M)
,
$
by
\begin{eqnarray}
\begin{array}{l}
\displaystyle 
  \left.
    \dd{\tau}{{X}^{k}(\vec{a},\tau)}
  \right|_{\tau=t}
  \left( \dd{x^{k}}{} \right)_{\vec{X}(\vec{a},t)}
  =
  \left( V_{i}^{k}(t) \dd{x^{k}}{} \right)_{\vec{X}(\vec{a},t)}
,
\\
\displaystyle
  \left.
    \dd{\tau}{{Y}^{k}(\vec{a},\tau)}
  \right|_{\tau=t}
  \left( \dd{x^{k}}{} \right)_{\vec{Y}(\vec{a},t)}
  =
  \left( V_{e}^{k}(t) \dd{x^{k}}{} \right)_{\vec{Y}(\vec{a},t)}
,
\end{array}
\hspace{1em}
\label{Eulerian Lagrangian velocity}
\end{eqnarray}
hereafter,
$
 \mathfrak{X}_{\Sigma}(M)
$
denotes the function space of
the divergence-free, tangent vector fields on $M$.
Mathematically,
the RHS's of these equations express the \textit{right translation}
of the vector field, $\bm{V}_{i}$ (resp. $\bm{V}_{e}$),
by the group operation $\vec{X}(t)$ (resp. $\vec{Y}(t)$).
As was discussed in \cite{araki2009comprehensive},
since the arguments of component function and basis do not agree with each other,
the LHS's of (\ref{Eulerian Lagrangian velocity}) 
are not proper differential topological objects;
the Lagrangian velocities in the RHS's form are appropriate 
for the calculus on manifolds.

In Lagrangian mechanics on Lie groups,
there exist two key mathematical structures:
the Lie bracket and the Riemannian metric.
The Lie bracket is necessary to determine the higher-order terms of the 
Taylor expansion of a composite function of PTMs.
The Riemannian metric is 
the inner product of two tangential vectors of PTMs
and defines the Lagrangian of the system.

Since the group operation of $G$ is defined by 
the compositions of function triplets
$
  (\vec{X}_{1},\vec{Y}_{1}) \circ (\vec{X}_{2},\vec{Y}_{2})
  =
  \big(\vec{X}_{1}(\vec{X}_{2}),\vec{Y}_{1}(\vec{Y}_{2})\big)
,
$
the Lie bracket of the associated Lie algebra
is given by
\begin{eqnarray}
&&\hspace{-2em}
 \LieBracket{\Big}{\GEV{V}{_{1}}}{\GEV{V}{_{2}}}
 =
 \Big(
  \nabla \times ( \Vi{1} \times \Vi{2} )
 ,
  \nabla \times ( \Ve{1} \times \Ve{2} )
 \Big)
,
\label{basic Lie bracket}
\end{eqnarray}
where 
$
 {\GEV{V}{_{k}}}
 =
 (\Vi{k},\Ve{k}) \in \mathfrak{g}
 =
 T_{e}G
 =
 \mathfrak{X}_{\Sigma}(M) \times \mathfrak{X}_{\Sigma}(M)
.
$
Since $M$ is three-dimensional and 
the vector fields considered here are divergence free,
the Lie bracket on 
$\mathfrak{X}_{\Sigma}(M)$
is given by $[\bm{a},\bm{b}]=\nabla\times(\bm{a}\times\bm{b})$.

In the present study,
the Riemannian metric at $(\vec{X},\vec{Y})\in G$
is defined by the combination of the integrals 
described by the Lagrangian and the Eulerian specifications
as follows:
\begin{eqnarray}
  \Braket{\Big}{\GEV{V}{_{1}}}{\GEV{V}{_{2}}}_{(\vec{X},\vec{Y})}
&&
  =
  \int_{\vec{a}\in M} {\rm{d}}^3{\vec{X}(\vec{a},t)}
    \big( \Vi{1}\cdot\Vi{2} \big)_{\vec{X}(\vec{a},t)}
 \nonumber\\&&
  +
  \frac{1}{\HALL^{2}}
  \int_{\vec{x}\in M} {\rm{d}}^3{\vec{x}} \Big\{
    \big[ (\nabla\times)^{-1} (\Vi{1}-\Ve{1}) \big]
 \nonumber\\&&\hspace{5em}
    \cdot
    \big[ (\nabla\times)^{-1} (\Vi{2}-\Ve{2}) \big]
  \Big\}_{\vec{x}}
,
\hspace{1em}
\label{def riemannian metric}
\end{eqnarray}
where 
${\rm{d}}^3{\vec{X}(\vec{a},t)}$
and $(\nabla\times)^{-1}$ are
the advected volume element at the time $t$ 
(which is initially located at $\vec{a}$)
and
the inverse of the curl operator,
respectively.
For practical calculations,
the first term is replaced by
$
  \int_{\vec{x}\in M}
  {\rm{d}}^3{\vec{x}}
  \big(
    \Vi{1}\cdot\Vi{2}
  \big)_{\vec{x}}
,
$
because the modulus of the volume element is conserved 
due to the incompressibility:
$
 |{\rm{d}}^3{\vec{X}(\vec{a},t)}|
 =
 |{\rm{d}}^3{\vec{a}}|
$
for all $t$.
Mathematically, this replacement implies 
the \textit{right invariance} of the Riemannian metric.

Since
the difference $\Vi{}-\Ve{}$ gives a current density $\HALL\bm{j}$,
the generated magnetic field is given by
$\bm{b} = \HALL^{-1} (\nabla\times)^{-1} (\Vi{1}-\Ve{1})$.
Thus,
the Riemannian metric expresses
the sum of 
the kinetic energy of the ion flow (with density $\rho=1$)
and
the magnetic field energy generated by the plasma current,
while the kinetic energy of the electron flow is assumed to be negligible.

In the present formulation,
the Riemannian metric coercively combines 
the two different vector spaces
by a subtraction operation,
although the implication of the operation is quite natural 
from a physical viewpoint.
In our previous study,
the coupling of two spaces is established by 
the group action of a semidirect product of two diffeomorphism groups,
while the Riemannian metric is defined by 
separately defined integrals.

A remark on the Lie algebraic structure should be made here.
Substituting 
\begin{eqnarray}
  \bm{V}_{i}=\bm{u}
,\ \ 
  \bm{V}_{e}=\bm{u}-\HALL\bm{j}
\label{Vi Ve to u j}
\end{eqnarray}
into (\ref{basic Lie bracket}) and (\ref{def riemannian metric}),
we obtain
the inner product of a $\GEV{V}{}$-variable and their Lie bracket
as follows:
\begin{eqnarray}
&&
  \Braket{\Big}{\GEV{V}{_{1}}}{
    \LieBracket{\big}{\GEV{V}{_{2}}}{\GEV{V}{_{3}}}
  }_{(e,e)}
  =
  \int_{\vec{x}\in M}{\rm{d}}^3\vec{x}\Big\{
    {\bm{u}_{1}}\cdot\big[\nabla\times(\bm{u}_{2}\times\bm{u}_{3})\big]
    \nonumber\\&&\hspace{2em}
    +
    \bm{b}_{1}
    \cdot
    \big(
      \bm{u}_{2}\times\bm{j}_{3}
      +
      \bm{j}_{2}\times\bm{u}_{3}
      -\HALL
      \bm{j}_{2}\times\bm{j}_{3}
    \big)
  \Big\}_{\vec{x}}
,
\end{eqnarray}
where $\bm{b}_{k}$ satisfies
$
  \bm{b}_{k}=(\nabla\times)^{-1}\bm{j}_{k}
,
$
$
  \nabla\cdot\bm{b}_{k}=0
.
$
The same integral can be obtained 
from the other Riemannian metric and the commutator,
i.e.,
from Eqs.(4) and (7) (or Eqs.(12) and (13)) of
our previous study \cite{Araki2015}.
This implies that
these two formulations provide
the same structure constant of the Lie algebra
if 
appropriate base functions 
such as the GEVs are applied,
and thus,
these two systems are equivalent
although their group structures are quite different from each other.
In other words,
these two formulations constitute
a kind of ``canonical transformation''
between the configuration spaces with different group structures.

\section{Variational calculation: derivation of the equation of 
motion and helicity conservation}

Action along a path 
$\gamma(t)=(\vec{X}(t),\vec{Y}(t))\in G$ ($t\in[0,1]$)
is given by
$
 S
 =
 \int_{0}^{1} L dt
,
$
where $L$ is the Lagrangian defined by
\begin{eqnarray}
 L=L({\gamma(t)},{\dot{\gamma}(t)})
 :=
 \frac{1}{2}
 \Braket{\big}{\GEV{V}{}}{\GEV{V}{}}_{\gamma(t)}
,
\label{action}
\end{eqnarray}
where
$
 \gamma(t+\tau)
 \approx
 (\exp(\tau\Vi{}(t)),\exp(\tau\Ve{}(t)))
 \circ \gamma(t)
$, 
$t\in[0,1]$, $\tau$ is a small parameter,
and
exp is the exponential map on $\mathfrak{X}_{\Sigma}(M)$.
Let
$
 \gamma(t;\delta)
$
be a perturbed path, 
where
$
 \gamma(t;0)=\gamma(t)
,
$
$$
 \gamma(t;\delta)
 \approx
 (\exp(\delta\bm{\xi}(t)),\exp(\delta\bm{\eta}(t)))
 \circ \gamma(t;0)
, 
$$
$\delta$ is a small parameter,
and
$
 (\bm{\xi}, \bm{\eta})\in \mathfrak{g} =
 \mathfrak{X}_{\Sigma}(M) \times \mathfrak{X}_{\Sigma}(M)
$
are the displacement fields.
%
%
Noticing that
the perturbation part of the velocity,
say $(\tilde{\Vi{}},\tilde{\Ve{}})$, obeys \textit{Lin constraints}
\begin{eqnarray}
 (\tilde{\Vi{}},\tilde{\Ve{}})
 =
 \partial_{t}\big({\bm{\xi}},{\bm{\eta}}\big)
 +
 \LieBracket{}{({\bm{\xi}},{\bm{\eta}})}{(\Vi{},\Ve{})}
\label{lin constraints}
\end{eqnarray}
(see Appendix \ref{Derivation of Lin constraints} for the derivation),
the first variation of the action is given by
\renewcommand{\dot}{{\partial_{t}}}
\begin{eqnarray}
 \delta S
 &=&
 \int_{0}^{1} dt \int_{\vec{x}\in M} {\rm{d}}^3{\vec{x}} \bigg\{
  \Vi{} \cdot \tilde{\Vi{}}
 \nonumber\\&&
  +
  \HALL^{-2}
  \Big[
   (\nabla\times)^{-1}
   (\Vi{}-\Ve{})
  \Big]
  \cdot
  \Big[
   (\nabla\times)^{-1}
   (\tilde{\Vi{}}-\tilde{\Ve{}})
  \Big]
 \bigg\}
\renewcommand{\sakujo}[1]{}\sakujo{
\nonumber\\
 &=&
 \int_{0}^{1} dt \int_{\vec{x}\in M} {\rm{d}}^3{\vec{x}} \bigg\{
  \Vi{} \cdot \Big(
   \dot{\bm{\xi}}+\nabla\times\big(\bm{\xi}\times\Vi{}\big)
  \Big)
 \nonumber\\&&
  +
  \HALL^{-2}
  \Big(
   \big(\nabla\times\big)^{-1}
   \big(\Vi{}-\Ve{}\big)
  \Big)
  \cdot
  \Big[
   \big(\nabla\times\big)^{-1}
   \big(
    \dot{\bm{\xi}}+\nabla\times(\bm{\xi}\times\Vi{})
    -
    \dot{\bm{\eta}}-\nabla\times(\bm{\eta}\times\Ve{})
   \big)
  \Big]
 \bigg\}
}
\nonumber\\
 &=&
 \int_{0}^{1} dt \int_{\vec{x}\in M} {\rm{d}}^3{\vec{x}} \bigg\{
  (\Vi{} + \bm{A}) \cdot \Big[
   \dot{\bm{\xi}}+\nabla\times\big(\bm{\xi}\times\Vi{}\big)
  \Big]
 \nonumber\\&&
  -
  \bm{A}
  \cdot
  \Big[
    \dot{\bm{\eta}}+\nabla\times\big(\bm{\eta}\times\Ve{}\big)
  \Big]
 \bigg\}
,
\label{first variation 1}
\end{eqnarray}
where $\bm{A}$ is the vector potential of the magnetic field
with Coulomb gauge divided by $\HALL$:
\begin{eqnarray}
 {\bm{A}}:=
  \HALL^{-2}
  \big(\nabla\times\big)^{-2}
  \big({\Vi{}}-{\Ve{}}\big)
  =
  \frac{\bm{a}}{\HALL}
.
\label{def of A}
\end{eqnarray}
%
%
In the present study,
we assume that the boundary integrals always vanish.
The expression in the second line yields the following two results:
first,
the conjugate momenta of $\Vi{}$ and $\Ve{}$ are given by
\begin{eqnarray}
\begin{array}{l}
\displaystyle\bigg.
  \Mi{}:=\frac{\delta L}{\delta\bm{V}_{i}}
  =
  \bm{V}_{i}
  +
  \bm{A}
  =
  \bm{u}+\frac{\bm{a}}{\HALL}
,
\\\displaystyle\bigg.
  \Me{}:=\frac{\delta L}{\delta\bm{V}_{e}}
  =
  - \bm{A}
  =
  - \frac{\bm{a}}{\HALL}
,
\end{array}
\label{conjugate momenta}
\end{eqnarray}
respectively;
second,
the variations due to $\bm{\xi}$ and $\bm{\eta}$ that satisfy
\begin{eqnarray}
\hspace{-1em}
  \dot{\bm{\xi}}+\nabla\times\big(\bm{\xi}\times\Vi{}\big)=\bm{0}
,
\hspace{1ex}
  \dot{\bm{\eta}}+\nabla\times\big(\bm{\eta}\times\Ve{}\big)=\bm{0}
\hspace{1em}
\label{relabeling symmetry}
\end{eqnarray}
retain the value of action.
In terms of Lin constraints (\ref{lin constraints}),
this reads as
$ (\tilde{\Vi{}},\tilde{\Ve{}}) = (\bm{0},\bm{0}) $,
i.e.,
the velocity fields along the perturbed paths are  
the same as those of the reference path.
This symmetry for the invariant action is well-known as
the \textit{particle relabeling symmetry}
\cite{PadhyeMorrison1996,Salmon1988};
%
%
we give a brief review in Appendix \ref{sec:relabeling symmetry}.

\vspace{1em}

By integration by parts of (\ref{first variation 1}) 
with respect to $t$ and $\vec{x}$
and 
changing the order of scalar triple products of vector fields,
we obtain
%
%
\renewcommand{\Mi}{({\Vi{}}+{\bm{A}})}%
\begin{eqnarray}
  \delta S
  &=&
  \int_{0}^{1} dt \int_{\vec{x}\in M} {\rm{d}}^3{\vec{x}} \bigg\{
    {\Mi{}} \cdot \partial_{t}{\bm{\xi}}
    +
   \nonumber\\&&\hspace{2em}
    \bm{\xi} \cdot
    \Big[
      \Vi{}\times\big(\nabla\times{\Mi{}}\big)
    \Big]
   \nonumber\\&&\hspace{2em}
    -
    {\bm{A}} \cdot \partial_{t}{\bm{\eta}}
    -
    \bm{\eta} \cdot \big[ \Ve{} \times (\nabla \times \bm{A}) \big]
  \bigg\}
,
\label{iLagrangeDirectProduct:eq:delta S process}
\\
  &=&
  \int_{\vec{x}\in M} {\rm{d}}^3{\vec{x}} \big(
    {\Mi{}} \cdot {\bm{\xi}}
    -
    {\bm{A}} \cdot {\bm{\eta}}
  \big)_{t=1}
 \nonumber\\&&
  -
  \int_{\vec{x}\in M} {\rm{d}}^3{\vec{x}} \big(
    {\Mi{}} \cdot {\bm{\xi}}
    -
    {\bm{A}} \cdot {\bm{\eta}}
  \big)_{t=0}
 \nonumber\\&&
  +
  \int_{0}^{1} dt \int_{\vec{x}\in M} {\rm{d}}^3{\vec{x}}
 \nonumber\\&&\hspace{1em}\times
  \bigg\{
    \bm{\xi} \cdot \Big[
      -\partial_t{\Mi{}}
      +
      \Vi{}\times\big(\nabla\times{\Mi{}}\big)
    \Big]
 \nonumber\\&&\hspace{6em}
    +
    \bm{\eta} \cdot \Big[
      \partial_t{\bm{A}}
      -
      \Ve{} \times (\nabla \times \bm{A})
    \Big]
 \bigg\}
.
\hspace{1em}
\label{iLagrangeDirectProduct:eq:delta S process 2}
\end{eqnarray}
%
%
Hamilton's principle, i.e.,
$\delta S=0$ for an \textit{arbitrary} perturbation
$(\bm{\xi}, \bm{\eta})$
with
fixed path end conditions,
leads to the following Euler-Lagrange equations:
\begin{eqnarray}
 \partial_t{\Mi{}}
 &=&
 \Vi{}\times\big(\nabla\times{\Mi{}}\big)
 -\nabla P_{i}
\label{iLagrangeDirectProduct:eq:EL dU/dt}
, \Big.
\\
 \partial_t\bm{A}
 &=&
 \Ve{} \times (\nabla \times \bm{A})
 -\nabla P_{e}
\label{iLagrangeDirectProduct:eq:EL dA/dt}
,
\end{eqnarray}
where $P_{i}$ and $P_{e}$ are 
the generalized pressures for each fluid.
Substituting (\ref{Vi Ve to u j}) and (\ref{def of A}),
and 
carrying out some calculations, 
we obtain
the evolution equation 
(\ref{eq of mot (u,b)}).
Note that,
in the limit $\HALL\to0$,
we obtain the standard MHD equations,
although 
the variable $\bm{A}=\bm{a}/\HALL$ diverges at $O(\HALL^{-1})$.

\vspace{1em}

Next,
we consider \textit{specific} perturbations 
$(\bm{\xi},\bm{\eta})$ that 
leave the value of action unchanged.
The
combinations of 
$\bm{\xi}=\Vi{}$ or $\nabla\times(\Vi{}+\bm{A})$
and
$\bm{\eta}=\Ve{}$ or $\nabla\times\bm{A}$
are
the candidates for the invariant action,
because
they cancel
the cubic terms of (\ref{iLagrangeDirectProduct:eq:delta S process}).
It is easy to see that
the conservation of total energy (\ref{total energy}) is obtained by setting
$
 (\bm{\xi}(t),\bm{\eta}(t))
 =
 (\Vi{},\Ve{})
,
$
which implies that 
the variation is taken in the direction of the path,
i.e.,
the variation is associated with the time translation.

By setting
$
 (\bm{\xi}(t),\bm{\eta}(t))
 =
 (\bm{0},\nabla\times\bm{A}(t))
,
$
the first variation of action 
(\ref{iLagrangeDirectProduct:eq:delta S process})
becomes
\begin{eqnarray}
 \delta S
 &=&
 \int_{0}^{1} dt
 \int_{\vec{x}\in M} {\rm{d}}^3{\vec{x}}
 \Big(-
  {\bm{A}} \cdot
  \big(\nabla\times{\dot{\bm{A}}}\big)
 \Big)
,
\nonumber\\
 &=&
 -
 \int_{0}^{1} dt
 \int_{\vec{x}\in M} {\rm{d}}^3{\vec{x}}
 \Big(
  \big(\nabla\times{{\bm{A}}}\big)
  \cdot
  \dot{\bm{A}}
 \Big)
,
\label{substitute (0,curlA)}
\end{eqnarray}
where 
the second line is obtained by
the integration by parts with respect to $\vec{x}$.
If the path $\gamma(t)$ satisfies the Euler-Lagrange equation 
(\ref{iLagrangeDirectProduct:eq:EL dA/dt}),
substitution of the RHS of (\ref{iLagrangeDirectProduct:eq:EL dA/dt})
into (\ref{substitute (0,curlA)}),
yields the vanishing first variation;
i.e.,
$\delta S=0$.
Noticing that the identity
\begin{eqnarray}
&&
 \int_{\vec{x}\in M} {\rm{d}}^3{\vec{x}}
 \Big(
  {\bm{A}} \cdot
  \big(\nabla\times{\dot{\bm{A}}}\big)
 \Big)
\nonumber\\&&\hspace{2em}
 =
 \frac{1}{2}
 \int_{\vec{x}\in M} {\rm{d}}^3{\vec{x}}
 \dd{t}{}
 \Big(
  (\nabla\times{\bm{A}}) \cdot \bm{A}
 \Big)
\nonumber
\end{eqnarray}
holds,
the integration by parts with respect to $t$ 
without fixed path end conditions 
results in
the conservation of magnetic helicity 
(\ref{magnetic helicity})
by Noether's first theorem:
\begin{eqnarray}
\delta S
 & = & 
 \frac{1}{2}
 \int_{0}^{1} dt
 \int_{\vec{x}\in M} {\rm{d}}^3{\vec{x}}
 \dd{t}{}
 \Big(-
  {\bm{A}} \cdot
  \big(\nabla\times{{\bm{A}}}\big)
 \Big)
\nonumber\\
 & = &
 \int_{\vec{x}\in M} {\rm{d}}^3{\vec{x}}
 \Big(
  {\bm{A}} \cdot
  \big(\nabla\times{\bm{A}}\big)
 \Big)_{t=0}
 \nonumber\\&&
 -
 \int_{\vec{x}\in M} {\rm{d}}^3{\vec{x}}
 \Big(
  {\bm{A}} \cdot
  \big(\nabla\times{\bm{A}}\big)
 \Big)_{t=1}
\nonumber\\
 & = &
 \HALL^{-2} H_{M}(1) - \HALL^{-2} H_{M}(0) =
 0
.
\end{eqnarray}
Similarly,
by setting
$
 (\bm{\xi}(t),\bm{\eta}(t))
 =
 (\nabla\times(\Vi{}(t)+\bm{A}(t)),\bm{0})
,
$
we obtain the conservation of hybrid helicity 
(\ref{hybrid helicity}) 
by Noether's first theorem:
\begin{eqnarray}
 \delta S
 &=&
 \int_{\vec{x}\in M} {\rm{d}}^3{\vec{x}}
 \Big[
  {\Mi{}}
  \cdot
  \big(\nabla\times{\Mi{}}\big)
 \Big]_{t=1}
 \nonumber\\&&
 -
 \int_{\vec{x}\in M} {\rm{d}}^3{\vec{x}}
 \Big[
  {\Mi{}} \cdot
  \big(\nabla\times{\Mi{}}\big)
 \Big]_{t=0}
\nonumber\\
 &=&
 \HALL^{-2} H_{H}(1) - \HALL^{-2} H_{H}(0) =
 0
.
\end{eqnarray}
In summary,
if the path $\gamma(t)$ locally satisfies 
the Euler-Lagrange equations 
(\ref{iLagrangeDirectProduct:eq:EL dU/dt}) and
(\ref{iLagrangeDirectProduct:eq:EL dA/dt}),
the combinations of perturbations
($\bm{\xi}$,$\bm{\eta}$)
=
$C_{E}$($\Vi{}$,$\Ve{}$)
and
($C_{i}\nabla\times{\Mi{}}$,$-C_{e}\nabla\times\bm{A}$)
retain
the value of the action,
where
$C_{E}$, $C_{i}$ and $C_{e}$ are arbitrary constants.
The derivation process clearly shows that
the magnetic (resp. hybrid) helicity is obtained 
by varying the integral path only 
on the $\Ve{}$-(resp. $\Vi{}$-)side of the configuration space;
i.e.,
the magnetic and hybrid helicities are obtained 
by the relabeling of $\Ve{}$ and $\Vi{}$.

\section{Differential topological description}

\renewcommand{\Mi}[1]{\bm{M}_{i#1}}%

In order to obtain mathematical insight to these conservation laws,
we revisit the discussion above 
using the differential topological expressions.

As a starting point, we notice that
the Lie bracket is also expressed by the Lie derivative of a vector field;
i.e.,
\begin{eqnarray}
  [\bm{\xi},\bm{\eta}]
  =
  L_{\bm{\eta}}\bm{\xi}
  =
  ({\eta^{j}}\dd{x^{j}}{\xi^{i}}-{\xi^{j}}\dd{x^{j}}{\eta^{i}})\dd{x^{i}}{}
\label{Lie bracket and Lie derivative}
\end{eqnarray}
for $\bm{\xi}$, $\bm{\eta} \in \mathfrak{X}(M)$.
Using this relation,
the first variation (\ref{first variation 1})
can be rewritten as
\begin{eqnarray}
&&
  \delta S
  =
  \int_{0}^{1}dt \Big\{
    \ParKet{\big}{\U{\Mi{}}}{(\partial_{t}+L_{\Vi{}}){\bm{\xi}}}+
    \ParKet{\big}{\U{\Me{}}}{(\partial_{t}+L_{\Ve{}}){\bm{\eta}}}
  \Big\}
,
\nonumber\\&&
\label{1st variation pairing}
\end{eqnarray}
where 
the underline denotes a differential 1-form
and 
the parenthesis
is the inner product between a differential 1-form and a vector field;
$
  \ParKet{}{*}{*}:
  \Omega_{1}(M)\times\mathfrak{X}(M)\to\mathbb{R}
,
$
where
$
  \ParKet{\big}{\bm{A}}{\bm{B}}
  =
   \int_{\vec{x}\in M} A_{i}B^{i}{\rm{d}}^3{\vec{x}}
,
$
$
  \bm{A}=A_{i}dx^{i}
  \in \Omega_{1}(M)
,
$
and
$
  \bm{B}=B^{i}\dd{x^{i}}{}
  \in \mathfrak{X}(M)
.
$
Integration by parts yields
\begin{eqnarray}
&&
  \delta S
  =
 \nonumber\\&&
  \Big\{
    \ParKet{\big}{\U{\Mi{}}}{{\bm{\xi}}}+
    \ParKet{\big}{\U{\Me{}}}{{\bm{\eta}}}
  \Big\}_{t=1}
  -
  \Big\{
    \ParKet{\big}{\U{\Mi{}}}{{\bm{\xi}}}+
    \ParKet{\big}{\U{\Me{}}}{{\bm{\eta}}}
  \Big\}_{t=0}
 \hspace{1em}\nonumber\\&&
  -
  \int_{0}^{1}dt \Big\{
    \ParKet{\big}{(\partial_{t}+L_{\Vi{}})\U{\Mi{}}}{{\bm{\xi}}}+
    \ParKet{\big}{(\partial_{t}+L_{\Ve{}})\U{\Me{}}}{{\bm{\eta}}}
  \Big\}
,
\label{1st var action in diff top}
\end{eqnarray}
which corresponds to (\ref{iLagrangeDirectProduct:eq:delta S process 2}).
Under the fixed path end conditions, 
$\bm{\xi}=\bm{\eta}=\bm{0}$ for $t=0$ and 1,
we obtain the Euler-Lagrange equation
\footnote{%
The Lie derivative of a differential 1-form is given by
$
  L_{\bm{\xi}}{\bm{\eta}}
  =
  \left(
    \xi^{k}\dd{x^k}{\eta_{j}}
    +
    {\eta_{k}}\dd{x^j}{\xi^{k}}
  \right) dx^{j}
$
for
$
  {\bm{\eta}}=\eta_{j} dx^{j} \in\Omega_{1}(M)
.
$
In vector analysis notation,
$
  L_{\bm{\xi}}{\bm{\eta}}
  =
  -\bm{\xi}\times(\nabla\times\bm{\eta})+\nabla(\bm\xi\cdot\bm\eta)
$
}%
\begin{eqnarray}
  {(\partial_{t}+L_{\Vi{}})\U{\Mi{}}} = -dP_{i}^\prime
,
\hspace{1ex}
  {(\partial_{t}+L_{\Ve{}})\U{\Me{}}} = -dP_{e}^\prime
,
\hspace{1em}
\label{eq of mot 1-form}
\end{eqnarray}
where $P_{i}^\prime$ and $P_{e}^\prime$ are introduced 
to satisfy the divergence-free condition.
These are the differential topological expressions of
(\ref{iLagrangeDirectProduct:eq:EL dU/dt}) and
(\ref{iLagrangeDirectProduct:eq:EL dA/dt}).
Since the exterior differentiation, $d$, is commutative with 
the Lie derivative of differential form
\cite{TurYanovsky1993,araki2009comprehensive},
using $dd=0$, we obtain
the exterior derivative of (\ref{eq of mot 1-form}) as follows:
\begin{eqnarray}
  (\partial_{t}+L_{\Vi{}})d\U{\Mi{}}=0
,
\hspace{1ex}
  (\partial_{t}+L_{\Ve{}})d\U{\Me{}}=0
.
\hspace{1em}
\label{eq of mot 2-form}
\end{eqnarray}
Since we consider three dimensional space and 
divergence-free vector fields and differential forms here,
there is a natural correspondence
between the differential 2-form and the vector field %
\footnote{%
The Lie derivative of a differential 2-form 
on a three-dimensional manifold
is given by
$
  L_{\bm{\xi}}\bm{\eta}
  =
  \left(
    \xi^{k}\dd{x^{k}}{\eta^{j}}
    -
    {\eta^{k}}\dd{x^{k}}{\xi^{j}}
    +
    {\eta^{j}}\dd{x^{k}}{\xi^{k}}
  \right) \epsilon_{jlm} dx^{l} \wedge dx^{m}
,
$
for 
$
  \bm{\eta}=\epsilon_{jlm} {\eta^{j}} dx^{l} \wedge dx^{m}
  \in\Omega_{2}(M)
,
$
where $\epsilon_{jlm}$ is the Levi-Civita symbol.
Due to the divergence-free condition,
transformation rule of the coefficients 
against the change of the local coordinate system
is given by
$
  \eta^{\prime j}(\vec{y})=
  \eta^i(\vec{x})\dd{x^i}{y^j}
$
for both the vector field and differential 2-form.
}.
%
%
Here,
we introduce the mapping 
$[*]:\Omega_{2}(M)\to\mathfrak{X}(M)$,
defined by 
\begin{eqnarray}
  \ParKet{\big}{\U{\bm{A}}}{\big[{\U{\U{\bm{\xi}}}}\big]}
  :=
  \int {\U{\bm{A}}} \wedge {\U{\U{\bm{\xi}}}}
,
\label{def 2-form to vec fld}
\end{eqnarray}
where
$
  {\U{\bm{A}}} \in \Omega_{1}(M)
,
$
and
$
  {\U{\U{\bm{\xi}}}} \in \Omega_{2}(M)
.
$
Using 
the relations 
$
  \big[L_{\Vx{}}d\U{\Mx{}}\big]
  =
  L_{\Vx{}}\big[d\U{\Mx{}}\big]
,
$
which are guaranteed by the divergence-free condition,
we obtain
\begin{eqnarray}
  (\partial_{t}+L_{\Vi{}})[d\U{\Mi{}}]=0
,
\hspace{1em}
  (\partial_{t}+L_{\Ve{}})[d\U{\Me{}}]=0
.
\label{eq of mot vec fld}
\end{eqnarray}
These equations obey
the invariant action conditions (\ref{relabeling symmetry}).
Thus,
by substituting 
$
  \bm{\xi} = \HALL C_i [d\U{\Mi{}}]
  = C_{i} (\HALL\nabla\times\bm{u}+\bm{b})
$ 
and
$
  \bm{\eta} = \HALL C_e [d\U{\Me{}}] = - C_{e} \bm{b}
$
into (\ref{1st var action in diff top}), 
we obtain the general helicity conservation law 
$H(1)-H(0)=0$
as a consequence of invariant action, 
where the constant $H$, 
which we call the mixed helicity hereafter, is given by
\begin{eqnarray}
  H
  & = &
  \HALL C_{i}\ParKet{\big}{\U{\Mi{}}}{[d\U{\Mi{}}]}
  +
  \HALL C_{e}\ParKet{\big}{\U{\Me{}}}{[d\U{\Me{}}]}
,
\label{mixed helicity}
\end{eqnarray}
and $C_{i}$ and $C_{e}$ are arbitrary constants.
Using (\ref{def 2-form to vec fld}),
$H$ is also written as the integral of the wedge product of 
differential forms:
\begin{eqnarray}
  H
  & = &
  \HALL \int \Big( C_{i}\U{\Mi{}} \wedge d\U{\Mi{}}
    + C_{e}\U{\Me{}} \wedge d\U{\Me{}} \Big)
.
\end{eqnarray}
This construction procedure is also regarded as an extension of 
the general helicity conservation laws found by Khesin and Chekanov 
\cite{KhesinChekanov1989}
to a direct product group case.

Since 
the exterior derivative of a differential 1-form 
on a three dimensional manifold
is given by the curl operation,
substituting (\ref{conjugate momenta})
into the mixed helicity (\ref{mixed helicity}),
we obtain each part of the mixed helicity as follows:
\begin{eqnarray}
  \ParKet{\big}{\U{\Mi{}}}{[d\U{\Mi{}}]}
  & = & 
  \frac{1}{\HALL^2}
  \int_{\vec{x}\in M}
    (\HALL\bm{u}+\bm{a}) \cdot \big(\HALL\nabla\times\bm{u}+\bm{b}\big)
  {\rm{d}}^3{\vec{x}}
,
\nonumber\\
  \ParKet{\big}{\U{\Me{}}}{[d\U{\Me{}}]}
  & = & 
  \frac{1}{\HALL^2}
  \int_{\vec{x}\in M}
    \bm{a} \cdot \bm{b}\,
  {\rm{d}}^3{\vec{x}}
.
\nonumber
\end{eqnarray}
Note that,
in the standard MHD limit $\HALL\to0$,
an asymptotic relation
$
  \Mi{}=\bm{V}_{i}+\bm{A}
  \approx
  \Me{}=\bm{A}=\bm{a}/\HALL
  \sim O(\HALL^{-1})
$
holds,
and thus
the hybrid helicity comes close to the magnetic helicity
as an $O(\HALL^{-2})$ quantity.
However,
when $C_{i}=-C_{e}$,
the singularity order of $H$ is reduced by $\HALL$,
i.e.,
the leading orders of these helicities cancel each other,
and
the mixed helicity becomes 
\begin{eqnarray}
  H=C_{i}
  \int {\rm{d}}^3{\vec{x}}\, \big[
    2 \bm{u}\cdot\bm{b} + \HALL \bm{u}\cdot(\nabla\times\bm{u})
  \big]
,
\end{eqnarray}
which converges to a finite value in the standard MHD limit.
Thus,
the conservation of cross helicity 
is shown to be a special case of 
the general conservation law for the mixed helicity.

\section{Double Beltrami function expansion}

\begin{figure}
\centering
\includegraphics[width=200pt]{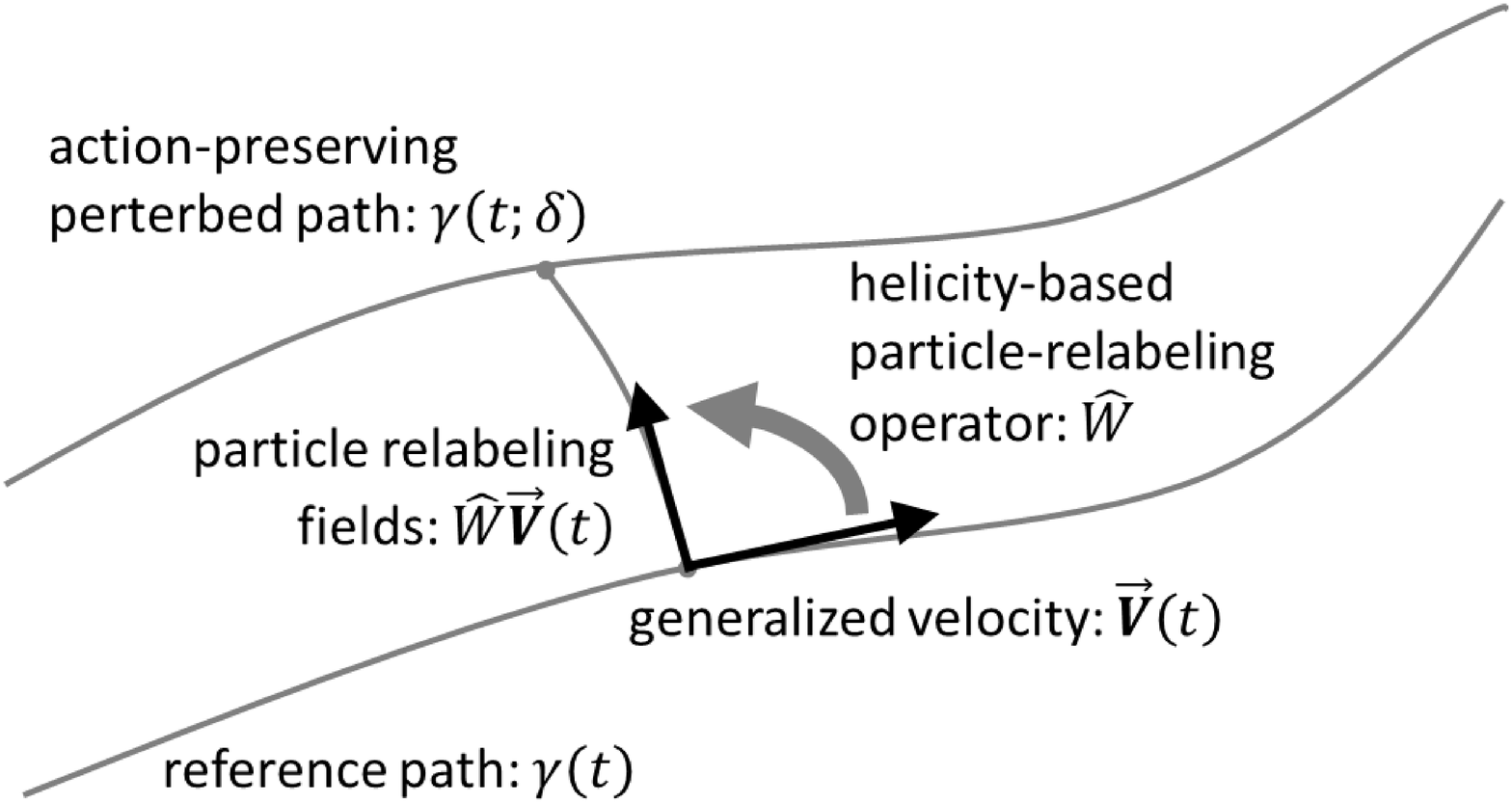}
\caption{\label{relabeling operator}
Relation among 
generalized velocity, $\vec{\bm{V}}$,
helicity-based, particle-relabeling operator, $\hat{W}$,
and action-preserving path variation,
$\gamma(t)\to\gamma(t;\delta)$.
}
\end{figure}

As is shown in the previous section,
the pair of vector fields 
($\HALL C_{i}{[d\U{\Mi{}}]}$, $\HALL C_{e}{[d\U{\Me{}}]}$)
satisfies the particle relabeling symmetry conditions 
(\ref{relabeling symmetry}).
This implies that
the integral path in the configuration space shifted in this direction 
retains the value of the action%
.
Noticing that
the vector fields pair is obtained by operation of appropriate operator 
on the ion and electron velocities pair as follows:
\begin{eqnarray}
  \left(\begin{array}{c}
    \HALL C_{i}{[d\U{\Mi{}}]} \Big.
  \\
    \HALL C_{e}{[d\U{\Me{}}]} \Big.
  \end{array}\right)
  =
  \hat{W}
  \left(\begin{array}{c}
    {\Vi{}} \Big.
  \\
    {\Ve{}} \Big.
  \end{array}\right)
,
\end{eqnarray}
where the integro-differential operator, $\hat{W}$, is defined by
\begin{eqnarray}
  \hat{W}
  :=
  \left(\begin{array}{cc}
    C_{i}(\HALL\nabla\times)^{-1}+C_{i}\HALL\nabla\times
  &
    -C_{i}(\HALL\nabla\times)^{-1}
  \\
    -C_{e}(\HALL\nabla\times)^{-1} & C_{e}(\HALL\nabla\times)^{-1}
  \end{array}\right)
,
\hspace{1em}
\label{CBPRop}
\end{eqnarray}
we recognize that
\textit{
the operation of\, $\hat{W}$ on the integral path of the action
physically implies infinitesimal particle relabeling operation}
(see Fig.\ref{relabeling operator}).
Thus,
we call $\hat{W}$ \textit{helicity-based, particle-relabeling operator} hereafter.

It seems reasonable to consider the eigenvalue problem of the operator $\hat{W}$,
because it is expected that
the spectral expansion by such eigenfunctions
should have some ``good'' properties for the description of 
the basic formulas and equations.
In the following, we will solve the eigenvalue problem, 
demonstrate the mode expansions of various quantities and equations,
and discuss the relation to 
the uniform background magnetic field effect
and 
the standard MHD limit.

The eigenvalue problem of the operator,
\begin{eqnarray}
  \hat{W} \GEV{V}{}
  =
  \EIGK \GEV{V}{}
,
\label{DBF problem}
\end{eqnarray}
is equivalent to the \textit{double Beltrami flow} (DBF) problem, 
which is given by the following coupled partial differential equations 
\cite{MahajanYoshida1998};
\begin{eqnarray}
  \HALL\nabla\times\bm{u}+\bm{b}=\frac{\EIGK}{C_{i}}\bm{u}
,
&\hspace{1em}&
  \bm{b}=-\frac{\EIGK}{C_{e}}(\bm{u}-\HALL\bm{j})
.
\label{double beltrami}
\end{eqnarray}

Note that
the eigenfunction of the DBF problem 
is constructed 
using a \textit{Beltrami flow}, 
i.e.,
the eigenfunctions of the curl operator,
say $\CHW{\psi}{(\vec{K},\HEL)}$, 
which satisfies
$$
  \nabla\times\CHW{\psi}{(\vec{K},\HEL)}
  =
  \HEL K \CHW{\psi}{(\vec{K},\HEL)}
,
$$
where
$\vec{K}$, $K>0$, and $\HEL=\pm1$ 
are the mode index, the associated eivenvalue, and 
the helicity of vector field,
respectively %
\footnote{%
Since $(\nabla\times)^2=-\triangle$ for divergence-free vector fields,
if the function is not harmonic,
the value $K>0$ is determined by 
the eigenvalue of the Laplacian:
$\triangle\bm{\psi}=-K^2\bm{\psi}$.
}.
The \textit{Chandrasekhar-Kendall function} 
on a cylindrical configuration
\cite{ChandrasekharKendall1957}
and
the \textit{complex helical waves}
on a periodic box or a Euclidean space
\cite{Waleffe1992}
are known as examples of the Beltrami flows.
Expanding the variables using $\CHW{\psi}{(\vec{K},\HEL)}$,
the operator $\hat{W}$ is reduced to a 2$\times$2 matrix as
$$
  \HEL
  \left(\begin{array}{cc}
   \displaystyle\bigg.
    \frac{C_{i}}{\HALL K}+C_{i}\HALL K
   &\displaystyle
    -\frac{C_{i}}{\HALL K}
   \\\displaystyle\bigg.
    -\frac{C_{e}}{\HALL K}
   &\displaystyle
    \frac{C_{e}}{\HALL K}
  \end{array}\right)
,
$$
for each expansion mode.
The eigenvalue of this matrix is given by
\begin{eqnarray}
  \EIGK(K,\HEL,\POL)
  & = & 
  \frac{\HEL}{2}\bigg\{
    \frac{C_{i}+C_{e}}{\HALL K}
    +
    C_{i}\HALL K
   \nonumber\\&&
    +
    \POL
    \Big[
      \big(
        \frac{C_{i}+C_{e}}{\HALL K}-C_{i}\HALL K
      \big)^2
      +
      4C_{i}^2
    \Big]^{\frac12}
  \bigg\}
,
\hspace{1em}
\label{eigenval CW}
\end{eqnarray}
where $\POL=\pm1$ is the polarity.
Note that, the eigenvalues for the assigned $K$ and $\HEL$ satisfy
\begin{eqnarray}
  \EIGK(K,\HEL,+) \EIGK(K,\HEL,-)=C_{i}C_{e},
\end{eqnarray}
and, in the standard MHD limit $\HALL\to0$, they become 
\begin{eqnarray}
  \EIGK(K,\HEL,+) \to \infty,\ 
  \EIGK(K,\HEL,-) \to 0
  &\mbox{\ \ for\ \ }&
  C_{i}\ne -C_{e}
, \ \ 
\\
  \EIGK(K,\HEL,\POL) \to \HEL \POL |C_{i}|
  &\mbox{\ \ for\ \ }&
  C_{i} = -C_{e}
. \ \ 
\end{eqnarray}

The eigenfunction of $\hat{W}$, say
$
  {\GEV{\Psi}{}}={}^{t}({\bm{\Psi}{_{i}}},{\bm{\Psi}{_{e}}})
,
$
is given by
\begin{eqnarray}
  {\GEV{\Psi}{}}{(\vec{K},\HEL,\POL)}
  &=&
  \left(\begin{array}{c}
    \left(
      \frac{1}{{\EIGK}(\vec{K},\HEL,\POL)}-\frac{\HALL K}{\HEL C_{e}}
    \right)
    \CHW{\psi}{(\vec{K},\HEL)}
  \\
    \frac{1}{{\EIGK}(\vec{K},\HEL,\POL)}\CHW{\psi}{(\vec{K},\HEL)}
  \end{array}\right)
,
\label{base functions V}
\end{eqnarray}
and we call the set of the eigenfunctions the DBF basis, hereafter.
If the Beltrami functions $\CHW{\psi}{}$ are orthonormal each other:
\begin{eqnarray}
  \int
    \CHW{\psi}{(\vec{K},\HEL_{K})}
    \cdot
    \overline{\CHW{\psi}{(\vec{P},\HEL_{P})}}\,
  {\rm{d}}^3\vec{x}
  &=&
  \delta_{\vec{K},\vec{P}}\,
  \delta_{\HEL_{K},\HEL_{P}}
,
\end{eqnarray}
hereafter, overline and $\delta$ 
denote complex conjugate and Kronecker's delta, respectively. 
The corresponding eigenfunctions $\GEV{\Psi}{}$ are orthogonal:
\newcommand{\DCM}{}%
\renewcommand{\DCM}[1]{\vec{#1},\HEL_{#1},\POL_{#1}}%
\renewcommand{\DCM}[1]{\widetilde{#1}}%
\begin{eqnarray}
  \BraKet{\Big}{ \GEV{\Psi}{}(\DCM{K}) }{
    \overline{\GEV{\Psi}{}(\DCM{P})}
  }
  &=&
  g(\DCM{K})\,
  \delta_{\vec{K},\vec{P}}\,
  \delta_{\HEL_{K},\HEL_{P}}\,
  \delta_{\POL_{K},\POL_{P}}
,
\hspace{2em}
\label{orthogonal V}
\end{eqnarray}
hereafter, 
the tilde denotes the set of mode indices, 
$\DCM{K}:=(\vec{K},\HEL_K,\POL_K)$, and
$g$ is the inner product of the base function 
${\GEV{\Psi}{}}{(\DCM{K}{})}$,
i.e., mathematically
the component of the Riemannian metric tensor
(\ref{def riemannian metric})
for the DBF basis, that is to say
\begin{eqnarray}
  g(\DCM{K})
  &:=&
  \left(\frac{1}{{\EIGK}(\DCM{K})}-\frac{\HALL K}{\HEL_{K} C_{e}}\right)^2
  +
  \frac{1}{C_{e}^2}
.
\label{coeff riemannian metric in Psi}
\end{eqnarray}
Substitution of $\bm{\Psi}{_{i}}$ and $\bm{\Psi}{_{e}}$ 
into (\ref{conjugate momenta})
yields the base functions of conjugate momenta space,
say
$
  \U{\GEV{\Psi}{}}={}^{t}(\U{\bm{\Psi}_{i}},\U{\bm{\Psi}_{e}})
,
$
as follows:
\begin{eqnarray}
  \U{\GEV{\Psi}{}}{(\DCM{K})}
  &=&
  \left(\begin{array}{c}
    \left( \frac{1}{{\EIGK}(\DCM{K})}
     -\frac{\HALL K}{\HEL C_{e}}-\frac{\HEL}{C_{e}\HALL K} \right)
    \CHW{\psi}{(\vec{K},\HEL)}
  \\
    \frac{\HEL}{C_{e}\HALL K}
    \CHW{\psi}{(\vec{K},\HEL)}
  \end{array}\right)
.
\hspace{1em}
\label{base functions M}
\end{eqnarray}
It is easy to check that the base functions
${\GEV{\Psi}{}}$ and $\U{\GEV{\Psi}{}}$
are bi-orthogonal each other; i.e.,
\begin{eqnarray}
  \ParKet{\big}{
    \U{\bm{\Psi}_{i}}{(\DCM{K})}
  }{
    {\bm{\Psi}_{i}}{(\DCM{P})}
  }
  +
  \ParKet{\big}{
    \U{\bm{\Psi}_{e}}{(\DCM{K})}
  }{
    {\bm{\Psi}_{e}}{(\DCM{P})}
  }
 \nonumber\\
  =
  g(\DCM{K})\,
  \delta_{\vec{K},\vec{P}}\,
  \delta_{\HEL_{K},\HEL_{P}}\,
  \delta_{\POL_{K},\POL_{P}}
.
\label{bi-orthogonal V and M}
\end{eqnarray}
Using these base functions,
we can rewrite the eigenvalue problem (\ref{DBF problem}) as follows:
\begin{eqnarray}
  \left(\begin{array}{c}
    \HALL C_{i} \curl\U{\bm{\Psi}_{i}}(\DCM{K})
  \\
    \HALL C_{e} \curl\U{\bm{\Psi}_{e}}(\DCM{K})
  \end{array}\right)
  =
  {\EIGK}(\DCM{K})
  \left(\begin{array}{c}
    {\bm{\Psi}_{i}}(\DCM{K})
  \\
    {\bm{\Psi}_{e}}(\DCM{K})
  \end{array}\right)
.
\label{Psi and underline Psi}
\end{eqnarray}
The generalized velocity and momentum are expanded 
using the eigenfunctions of $\hat{W}$
as
\begin{eqnarray}
  \GEV{V}{}
  =
  \sum_{\DCM{K}}
    \GEC{V}(\DCM{K})\,
    \GEV{\Psi}{}(\DCM{K})
,
  &\hspace{1em}&
  \GEV{M}{}
  =
  \sum_{\DCM{K}}
    \GEC{V}(\DCM{K})\,
    \U{\GEV{\Psi}{}}(\DCM{K})
,
\label{DBF expansion of V and M}
\end{eqnarray}
where the expansion coeffcient, $\GEC{V}(\DCM{K})$,
is obtained by the inner product,
\renewcommand{\DCM}[1]{\tilde{#1}}
\begin{eqnarray}
  \GEC{V}(\DCM{K})
  &=&
  \BraKet{\Big}{
    \GEV{V}{}
  }{\,
    \CC{\GEV{\Psi}{(\GEM{K}{})}}\,
  }
.
\end{eqnarray}
Since
$\Vi{}=\bm{u}$ and 
$\HALL\curl\U{\Me{}}=\HALL\curl(\bm{a}/\HALL)=-\bm{b}$,
the expansion coefficients, $\GEC{V}{(\vec{K},\HEL,\POL)}$, 
are determined by the following simultaneous equations:
\begin{eqnarray}
  \CHC{u}{(\vec{K},\HEL)}
  &=&
  \left(\frac{1}{{\EIGK}(K,\HEL,+)}-\frac{\HALL K}{\HEL C_{e}}\right)
  \GEC{V}{(\vec{K},\HEL,+)}
 \nonumber\\&&
  +
  \left(\frac{1}{{\EIGK}(K,\HEL,-)}-\frac{\HALL K}{\HEL C_{e}}\right)
  \GEC{V}{(\vec{K},\HEL,-)}
,
\nonumber
\\
  \CHC{b}{(\vec{K},\HEL)}
  &=&
  -\frac{1}{C_{e}}\left( \GEC{V}{(\vec{K},\HEL,+)} + \GEC{V}{(\vec{K},\HEL,-)} \right)
,
\label{determine V}
\end{eqnarray}
where $\CHC{u}{(\vec{K},\HEL)}$ and $\CHC{b}{(\vec{K},\HEL)}$ are 
the spectral expansion coefficients of the velocity and magnetic fields
with respect to the basis $\{\CHW{\psi}{(\vec{K},\HEL)}\}$,
respectively.
Unfortunately,
the expansion coefficients $\GEC{V}{}$ converge to zero or diverge 
in the limit $\HALL\to0$, unless $C_{i}=-C_{e}$.

The energy and the mixed helicity are obtained 
by substituting (\ref{DBF expansion of V and M})
into (\ref{action}) and (\ref{mixed helicity})
and using the relations (\ref{orthogonal V}), (\ref{bi-orthogonal V and M}),
and (\ref{Psi and underline Psi})%
; i.e.,
\begin{eqnarray}
  E
  &=&
  \frac12
  \BraKet{\Big}{
    \sum
    \GEC{V}(\DCM{K})
    {\GEV{\Psi}{}}(\DCM{K})
  }{
    \sum
    \GEC{V}(\DCM{P})
    {\GEV{\Psi}{}}(\DCM{P})
  }
\nonumber\\
  &=&
  \frac12
  \sum_{\DCM{K}}
  g(\DCM{K})
  \left|\GEC{V}{(\DCM{K})}\right|^2
,
\\
  H
  &=&
  \HALL C_{i}\ParKet{\Big}{
    \sum
    \GEC{V}(\DCM{K})
    \U{\bm{\Psi}_{i}}(\DCM{K})
  }{
    \sum
    \GEC{V}(\DCM{P})
    \curl\U{\bm{\Psi}_{i}}(\DCM{P})
  }
\nonumber\\&&
  +
  \HALL C_{e}\ParKet{\Big}{
    \sum
    \GEC{V}(\DCM{K})
    \U{\bm{\Psi}_{e}}(\DCM{K})
  }{
    \sum
    \GEC{V}(\DCM{P})
    \curl\U{\bm{\Psi}_{e}}(\DCM{P})
  }
\nonumber\\
  &=&
  \sum_{\DCM{K}}
  g(\DCM{K})
  \EIGK(\DCM{K})
  \left|\GEC{V}{(\DCM{K})}\right|^2
.
\end{eqnarray}
Applying the DBF expansion to the equation of motion yields 
the following simultaneous equations for the expansion coefficients,
${\GEC{V}{(\DCM{K}{};t)}}$:
\begin{eqnarray}
  \sdd{t}{}
  \CC{\GEC{V}{(\DCM{K}{};t)}}
  =
  g(\DCM{K}{})^{-1}
  \sum_{\DCM{P}{}}\sum_{\DCM{Q}{}}
 &&
  \tripleHMHD{\big}{\DCM{K}{}}{\DCM{P}{}}{\DCM{Q}{}}\,
  \EIGK(\DCM{Q}{})\,
 \nonumber\\&&\times
  \GEC{V}{(\DCM{P}{};t)}\,
  \GEC{V}{(\DCM{Q}{};t)}
,
\label{eq of mot in double Beltram coeff}
\end{eqnarray}
where the symbol
$
  \tripleHMHD{\big}{\DCM{K}{}}{\DCM{P}{}}{\DCM{Q}{}}
$
is given by
\begin{eqnarray}
  \tripleHMHD{\big}{\DCM{K}{}}{\DCM{P}{}}{\DCM{Q}{}}
  &&=
  \frac{1}{\HALL}
  \Bigg[
    \frac{1}{C_{i}}
    \bigg(\frac{1}{{\EIGK}(\DCM{K}{})}-\frac{\HALL K}{\HEL_{K} C_{e}}\bigg)
    \bigg(\frac{1}{{\EIGK}(\DCM{P}{})}-\frac{\HALL P}{\HEL_{P} C_{e}}\bigg)
   \nonumber\\&&\times
    \bigg(\frac{1}{{\EIGK}(\DCM{Q}{})}-\frac{\HALL Q}{\HEL_{Q} C_{e}}\bigg)
    +
    \frac{1}{
      C_{e}
      \EIGK(\DCM{K}{})
      \EIGK(\DCM{P}{})
      \EIGK(\DCM{Q}{})
    }
  \Bigg]
 \nonumber\\&&\times
  \int
    \CHW{\psi}{(\vec{K},\HEL_{K})}
    \cdot\left(
      \CHW{\psi}{(\vec{P},\HEL_{P})}
    \times
      \CHW{\psi}{(\vec{Q},\HEL_{Q})}
    \right)
  {\rm{d}}^3\vec{x}
.
\nonumber\\
\label{str. const. core DCB}
\end{eqnarray}
The derivations of 
(\ref{eq of mot in double Beltram coeff}) and (\ref{str. const. core DCB}) 
are summarized in Appendix \ref{derivation str const}.
Note that 
the symbol 
$
  \tripleHMHD{\big}{\DCM{K}{}}{\DCM{P}{}}{\DCM{Q}{}}
$
is skew-symmetric between two arbitrary argument sets,
because
the integrand is given by a scalar triple product of vector-valued functions
whereas
the coefficient is symmetric.
Due to this skew symmetry,
we can easily prove the conservation laws of 
the energy and the mixed helicity from the expression 
(\ref{eq of mot in double Beltram coeff}) as follows:
\begin{eqnarray}
  \sdd{t}{E}
  =
 &&
  \frac12
  \sum_{\DCM{K}{}}\sum_{\DCM{P}{}}\sum_{\DCM{Q}{}}
  \tripleHMHD{\big}{\DCM{K}{}}{\DCM{P}{}}{\DCM{Q}{}}\,
  \EIGK(\DCM{Q}{})\,
 \nonumber\\&&\hspace{1em}\times
  \GEC{V}{(\DCM{K}{};t)}\,
  \GEC{V}{(\DCM{P}{};t)}\,
  \GEC{V}{(\DCM{Q}{};t)}
  +
  c.c.
  =0
,
\hspace{1em}
\\\nonumber\\
  \sdd{t}{H}
  =
 &&
  \sum_{\DCM{K}{}}\sum_{\DCM{P}{}}\sum_{\DCM{Q}{}}
  \tripleHMHD{\big}{\DCM{K}{}}{\DCM{P}{}}{\DCM{Q}{}}\,
  \EIGK(\DCM{K}{})\,
  \EIGK(\DCM{Q}{})\,
 \nonumber\\&&\hspace{1em}\times
  \GEC{V}{(\DCM{K}{};t)}\,
  \GEC{V}{(\DCM{P}{};t)}\,
  \GEC{V}{(\DCM{Q}{};t)}
  +
  c.c.
  =0
.
\end{eqnarray}
Thus,
the DBFs, i.e., the eigenfunctions of the helicity-based particle-relabeling operator 
are shown to constitute a family of orthogonal function bases that yields 
a remarkably simple spectral representation of the equation of motion.
Especially,
the mixed helicity conservation is naturally built in this representation
due to the skew-symmetry of the coefficients of the quadratic terms.
In the previous study,
wherein the generalized {\Elsasser} variables (GEV) expansion
of the HMHD system was presented,
we conjectured that the conservation of the modified cross helicity
might reflect some symmetry intrinsic in the system \cite{Araki2015}.
Since the DBFs become the GEVs for $C_{i}=-C_{e}=1$,
the modified helicity conservation is now recognized as 
the consequence of a special case of the particle relabeling symmetry
for ion and electron flows.

\section{Consideration of the uniform background magnetic field}

For some practical applications,
it is important to consider
the influence of the uniform background magnetic field
on the dynamics of plasmas.

Here,
we mathematically consider 
the influence of a background magnetic field, say $\bm{B}_{0}$,
which is a harmonic function:
$\nabla\times\bm{B}_{0}=\bm{0}$, $\nabla\cdot\bm{B}_{0}={0}$.
Substitution of $\bm{b}+\bm{B}_{0}$ into
the equations of motion (\ref{eq of mot vec fld})
yields
\begin{eqnarray}
\begin{array}{l}
  (\partial_{t}+L_{\Vi{}})(\HALL[d\U{\Mi{}}]+\bm{B}_{0})=\bm{0}
,\Big.
\\
  (\partial_{t}+L_{\Ve{}})(\HALL[d\U{\Me{}}]-\bm{B}_{0})=\bm{0}
.\Big.
\end{array}
\end{eqnarray}
When the amplitudes of the variables are sufficiently small
compared to the modulus of $\bm{B}_{0}$,
these equations are approximated as
\begin{eqnarray}
  \HALL\partial_{t}\big[d\U{\bm{M}_{i}}\big]=L_{\bm{B}_{0}}{\Vi{}}
,
\hspace{1em}
&&
  \HALL\partial_{t}\big[d\U{\bm{M}_{e}}\big]=-L_{\bm{B}_{0}}{\Ve{}}
,
\end{eqnarray}
where the identity for two vector fields
$
  L_{\bm{\xi}}{\bm{\eta}}=-L_{\bm{\eta}}{\bm{\xi}}
$
is used
(see Eq.(\ref{Lie bracket and Lie derivative})).
These linear simultaneous equations can be described 
using the operator $\hat{W}$ as:
\begin{eqnarray}
&&
  \dd{t}{}
  \hat{W}
  \left(\begin{array}{c} {\bm{V}_{i}} \\ {\bm{V}_{e}} \end{array}\right)
  =
  L_{\bm{B}_{0}}
  \left(\begin{array}{c} {\bm{V}_{i }} \\ {\bm{V}_{e}} \end{array}\right)
.
\end{eqnarray}
Thus, 
the linear waves are shown to be expressed 
by the eigenfunctions of 
the DBF problem (\ref{base functions V}),
with $C_{i}=1$ and $C_{e}=-1$.
The eigenvalue become
$$
  \EIGK(\DCM{K})=
  \frac{\HEL}{2}( \HALL K + \POL \sqrt{ (\HALL K)^2 + 4 } )
.
$$
For typical velocity and magnetic field variables,
the linear simultaneous equations are given by
\begin{eqnarray}
\begin{array}{l}
 {\dot{\bm{u}}}
 =
 \bm{j}\times\bm{B}_{0}
 -
 \nabla P
,\Big.
\\
 {\dot{\bm{b}}}
 =
 \nabla\times\big(
  (\bm{u}-\HALL\nabla\times\bm{b})\times\bm{B}_{0}
 \big)
.\Big.
\end{array}
\label{linearized HMHD}
\end{eqnarray}
The eigenfunctions of the linearized HMHD system 
(\ref{linearized HMHD}) 
are known as the \textit{generalized {\Elsasser} variables} (GEV) 
\cite{Galtier2006}.
Physically,
the GEVs describe the ion cyclotron or whistler waves in plasmas.
The phase velocity for an assigned $\DCM{K}$ is 
$
  \omega(\DCM{K})
  := B_{0} k_{\parallel} \EIGK(\DCM{K})^{-1}
,
$
where $k_{\parallel}$ is the wavenumber of $\CHW{\psi}{(\vec{K},\HEL_{K})}$
in the direction of $\bm{B}_{0}$.
The equations of motion for this case are obtained by
substituting 
$
  \GEC{V}{(\DCM{K};t)}e^{-i\omega(\DCM{K})t}
$
into $\GEC{V}{(\DCM{K};t)}$.

It is interesting that
the GEV expansion coefficients of the basic variables 
are simply and hierarchically expressed by 
$\GEC{V}{}$ multiplied by the powers of $\EIGK$ 
as follows:
\begin{eqnarray}
\begin{array}{rcl}
  \Ve{}
  &=&
  \bm{u} - \HALL\nabla\times\bm{b}
\\
  &=&
  \sum_{\DCM{K}{}}
  {\EIGK}(\DCM{K}{})^{-1}\ 
  \GEC{V}{(\DCM{K}{})}\ 
  \CHW{\psi}{(\vec{K},\HEL_{K})}
,
\\
\\
  -\HALL[d\U{\Me{}}]
  &=&
  \bm{b}
\\
  &=&
  \sum_{\DCM{K}{}}
  \GEC{V}{(\DCM{K}{})}\ 
  \CHW{\psi}{(\vec{K},\HEL_{K})}
,
\\
\\
  \Vi{}
  &=&
  \bm{u}
\\
  &=&
  \sum_{\DCM{K}{}}
  {\EIGK}(\DCM{K}{})\ 
  \GEC{V}{(\DCM{K}{})}\ 
  \CHW{\psi}{(\vec{K},\HEL_{K})}
,
\\
\\
  \HALL[d\U{\Mi{}}]
  &=&
  \HALL\nabla\times\bm{u}+\bm{b}
\\
  &=&
  \sum_{\DCM{K}{}}
  {\EIGK}(\DCM{K}{})^2\ 
  \GEC{V}{(\DCM{K}{})}\ 
  \CHW{\psi}{(\vec{K},\HEL_{K})}
.
\end{array}
\end{eqnarray}

Note that
helicity parameters for the linear wave modes have
the following significant properties.

Firstly,
the eigenvalues (\ref{eigenval CW}) do not diverge 
in the standard MHD limit, $\HALL\to0$:
$$
  \EIGK(\DCM{K}{})
  \approx
  \HEL_{K} \POL_{K} + \frac{\HALL}{2} \HEL_{K} K + \frac{\HALL^2}{8} \HEL_{K} \POL_{K} K^2
  \longrightarrow
  \HEL_{K} \POL_{K} .
$$
This convergence 
leads to the finiteness of the following quantities in that limit:
coefficients of the base functions (\ref{base functions V}):
\begin{eqnarray}
  {\GEV{\Psi}{}}{(\DCM{K}{})}
  && \approx 
  \left(\begin{array}{r}
    (\HEL_{K} \POL_{K} + \frac{\HALL}{2}\HEL_{K} K)
    \CHW{\psi}{(\vec{K},\HEL_{K})}
  \\
    (\HEL_{K} \POL_{K} - \frac{\HALL}{2}\HEL_{K} K)
    \CHW{\psi}{(\vec{K},\HEL_{K})}
  \end{array}\right)
 \nonumber\\
  && \longrightarrow
  \HEL_{K} \POL_{K} 
  \left(\begin{array}{r}
    \CHW{\psi}{(\vec{K},\HEL_{K})}
  \\
    \CHW{\psi}{(\vec{K},\HEL_{K})}
  \end{array}\right)
,
\nonumber
\end{eqnarray}
the Riemannian metric (\ref{coeff riemannian metric in Psi}):
$$
  g(\DCM{K}{})
  =
  2+{\HALL}^{2}K^2
  \longrightarrow
  2
,
$$
the coefficient of the RHS of (\ref{str. const. core DCB}):
\begin{eqnarray}
&&
  \frac{1}{\HALL}
  \Bigg[
    \bigg(\frac{1}{{\EIGK}(\DCM{K}{})}+\frac{\HALL K}{\HEL_{K}}\bigg)
    \bigg(\frac{1}{{\EIGK}(\DCM{P}{})}+\frac{\HALL P}{\HEL_{P}}\bigg)
    \bigg(\frac{1}{{\EIGK}(\DCM{Q}{})}+\frac{\HALL Q}{\HEL_{Q}}\bigg)
   \nonumber\\&&\hspace{1em}
    -
    \frac{1}{
      \EIGK(\DCM{K}{})
      \EIGK(\DCM{P}{})
      \EIGK(\DCM{Q}{})
    }
  \Bigg]
\nonumber\\&&
  \approx
  \sigma_{K} \sigma_{P} \sigma_{Q} s_{K} s_{P} s_{Q} 
  \bigg(
    s_{K} K
  + s_{P} P
  + s_{Q} Q
  \bigg)
 \nonumber\\&&\hspace{1em}
  + \frac{1}{8} \HALL^2
  \sigma_{K} \sigma_{P} \sigma_{Q} K P Q
  \bigg(
      \frac{ s_{P} P + s_{Q} Q }{ s_{K} K }
 \nonumber\\&&\hspace{3em}
    + \frac{ s_{Q} Q + s_{K} K }{ s_{P} P }
    + \frac{ s_{K} K + s_{P} P }{ s_{Q} Q }
    + 2
  \bigg)
  + o(\HALL^2)
 \nonumber\\&&
  \longrightarrow
  \sigma_{K} \sigma_{P} \sigma_{Q} s_{K} s_{P} s_{Q} 
  \bigg(
    s_{K} K
  + s_{P} P
  + s_{Q} Q
  \bigg)
.
\end{eqnarray}
%
%
Since the simultaneous equations (\ref{determine V}) converge to
\begin{eqnarray}
&&
  \CHC{u}{(\vec{K},\HEL_{K})}
  =
  \HEL_{K} \POL_{K} 
  \GEC{V}{(\vec{K},\HEL_{K},-)}
  -
  \HEL_{K} \POL_{K} 
  \GEC{V}{(\vec{K},\HEL_{K},+)}
,
\nonumber\\&&
  \CHC{b}{(\vec{K},\HEL_{K})}
  =
  \GEC{V}{(\vec{K},\HEL_{K},+)} + \GEC{V}{(\vec{K},\HEL_{K},-)}
,
\end{eqnarray}
we obtain the MHD limit of the expansion coefficient, $\GEC{V}{}$:
\begin{eqnarray}
  \GEC{V}{(\DCM{K}{},\HEL_{K},\POL_{K})}
  =
  \CHC{b}{(\vec{K},\HEL_{K})}
  -
  \HEL_{K} \POL_{K} \,
  \CHC{u}{(\vec{K},\HEL_{K})}
,
\end{eqnarray}
and thus,
the equation of motion (\ref{eq of mot in double Beltram coeff})
has the standard MHD limit.
The coefficients, $\GEC{V}{}$,
are associated with 
the conventional {\Elsasser} variables by the formula
\begin{eqnarray}
\begin{array}{rcl}
  \bm{z}_{+}
  & = &
  \bm{u}+\bm{b}
\\
  & = &
  \GEC{V}{(K,+,-)}\CHW{\psi}{(K,+)}
  +
  \GEC{V}{(K,-,+)}\CHW{\psi}{(K,-)}
,
\\
  \bm{z}_{-}
  & = &
  \bm{u}-\bm{b}
\\
  & = &
  -
  \GEC{V}{(K,+,+)}\CHW{\psi}{(K,+)}
  -
  \GEC{V}{(K,-,-)}\CHW{\psi}{(K,-)}
.
\end{array}
\hspace{1em}
\end{eqnarray}
The base function in momentum space $\U{\GEV{\Psi}{}}$%
, on the other hand, 
diverges on the order of $\HALL^{-1}$,
at which the diverging $\bm{a}/\HALL$ term is reflected.

Secondly,
the singularity order of the mixed helicity (\ref{mixed helicity}) 
reduces by $\HALL$
and
the constant $H$ become the modified cross helicity,
which converges to the cross helicity:
\begin{eqnarray}
  H_{C}
&&
  =
  \int {\rm{d}}^3{\vec{x}}\, \big[
    2 \bm{u}\cdot\bm{b} + \HALL \bm{u}\cdot(\nabla\times\bm{u})
  \big]
\nonumber
\\&&
  \longrightarrow
  2 \int {\rm{d}}^3{\vec{x}}\
    \bm{u}\cdot\bm{b}
.
\end{eqnarray}
In our previous study, it was shown that
the conservation of the modified cross helicity
is naturally derived from the GEV representation of the HMHD dynamics.

\section{Discussion}

In the present study,
we considered the helicity conservation laws of HMHD system
from Lagrangian mechanical, invariant action theory viewpoint.
The hybrid and magnetic helicity conservation laws 
were derived as consequences of the particle relabeling symmetry 
of the ion and electron flows, respectively.
To prove the conservation laws,
it is convenient to use
the pair of ion and electron velocity fields 
(\ref{Eulerian Lagrangian velocity}) as basic variables,
while
that of fluid velocity and current fields had been used
in our previous study \cite{Araki2015}.
Mathematically,
this variables change was carried out by changing 
the configuration space of HMHD system
from semidirect product group to direct product one.

Furthermore,
associated integral path variation of the invariant action 
was shown to be expressed by the operation of 
the helicity-based, particle-relabeling operator (\ref{CBPRop})
on the reference path,
which maps the generalized velocities
$
 (\Vi{},\Ve{})
$ 
to the action-preserving, particle-relabeling fields
$(
  \HALL C_{i}{[d\U{\Mi{}}]}
  ,
  \HALL C_{e}{[d\U{\Me{}}]}
)$.

The eigenfunctions of the relabeling operator are DBFs,
which are well-known, force-free solutions of the HMHD system
\cite{MahajanYoshida1998},
and
found to provide a family of orthogonal function bases
that yields the spectral representation of the equation of motion
with a remarkably simple form.
Thus,
the GEV based formulation we had discussed in \cite{Araki2015}
is now understood as 
\textit{an example} of more wider class of orthogonal 
function expansion of the HMHD equations,
since the GEVs are special case of the DBFs ($C_{i}=-C_{e}=1$).

The implication of this eigenvalue problem
may be well-understood
by considering 
the correspondence between the Lagrangian and Hamiltonian mechanics.

It is well-known that
the Lie algebraic structure naturally induces
a so-called Lie-Poisson structure
on the dual space of the Lie algebra
by defining the Poisson bracket by
$
  \LieBrace{}{A}{B}(\mu)
  =
  \ParKet{\big}{\mu}{
    \LieBracket{}{
      \frac{\partial A}{\partial\mu}
    }{
      \frac{\partial B}{\partial\mu}
    }
  }
,
$
where
$
  \mu \in \mathfrak{g}^{*}
$
is an element of the generalized momentum space,
and
$
  A,\,B \in {\cal{F}}(\mathfrak{g}^{*})
$
are the functionals of the generalized momenta
\cite{MarsdenRatiu1994}.
In the incompressible HMHD case,
the Poisson bracket 
based on (\ref{basic Lie bracket}) and (\ref{def riemannian metric})
becomes
\begin{eqnarray}
  \LieBrace{\big}{A}{B}(\GEV{M}{})
  & = &
  \ParKet{\Big}{
    \Mi{}
  }{
    \nabla\times\big(
      \frac{\delta A}{\delta\Mi{}}
      \times
      \frac{\delta B}{\delta\Mi{}}
    \big)
  }
 \nonumber\\&&
  +
  \ParKet{\Big}{
    \Me{}
  }{
    \nabla\times\big(
      \frac{\delta A}{\delta\Me{}}
      \times
      \frac{\delta B}{\delta\Me{}}
    \big)
  }
.
\label{Poisson bracket}
\end{eqnarray}
When $B$ is the Hamiltonian obtained by the Legendre transformation 
of the Lagrangian (\ref{action}),
which results in $B(\GEV{M}{})=L(\GEV{V}{})$,
the functional derivatives of $B$ are given by
$
  \frac{\delta B}{\delta\Mi{}}=\Vi{}
,
$
$
  \frac{\delta B}{\delta\Me{}}=\Ve{}
.
$
Integration by parts of (\ref{Poisson bracket}) yields
\begin{eqnarray}
  \LieBrace{\big}{A}{B}(\GEV{M}{})
  & = &
  \int{\rm{d}}^3\vec{x}\bigg[
    \frac{\delta A}{\delta\Mi{}}
    \cdot
    \Big(
      \Vi{}
      \times
      \big(
        \nabla\times\Mi{}
      \big)
    \Big)
 \nonumber\\&&\hspace{3em}
  +
    \frac{\delta A}{\delta\Me{}}
    \cdot
    \Big(
      \Ve{}
      \times
      \big(
        \nabla\times\Me{}
      \big)
    \Big)
  \bigg]
.\hspace{1em}
\end{eqnarray}
By setting the derivatives 
$
  \frac{\delta A}{\delta\Mi{}}=\bm{\xi}
,
$
$
  \frac{\delta A}{\delta\Me{}}=\bm{\eta}
,
$
the cubic terms of the first variation
(\ref{iLagrangeDirectProduct:eq:delta S process})
are reproduced,
and thus,
the action-preserving variation 
is shown to correspond to
the functional derivative of a certain Casimir function.

In the Hamiltonian mechanical approach to 
the stability problem of the equilibrium solutions,
the DBF are known to constitute 
the dynamically accessible variations
that \textit{a priori} satisfy the conservation laws for energy and Casimirs
\cite{HirotaETAL2006}.
In the incompressible HMHD case,
the Casimirs are given by the magnetic and hybrid helicities.
In the Lagrangian mechanical approach, 
on the other hand,
they are obtained from the invariant action.
Thus,
the eigenvalue problem for the invariant action is naturally 
described as the DBF problem.

Since the DBFs for assigned $C_{i}$ and $C_{e}$ 
were orthogonal each other,
by using them as base functions
we could obtain a general form of the ``normal mode'' expansion of 
the Riemannian metric, the structure constants of the Lie algebra,
and the equation of motion.
The combinations of $C_{i}$ and $C_{e}$ are arbitrary,
i.e.,
the DBF basis has two degrees of freedom.
By changing the values of $C_{i}$ and $C_{e}$,
we obtained a family of ``canonical'' transformations
between the spectral representations of the equation of motion.
The spectral representations of the equation of motion
formally have a common mathematical expression given by 
(\ref{eq of mot in double Beltram coeff}),
which is known as the \textit{Euler-Poincare equation}
(Chapter 13 of Ref.\cite{MarsdenRatiu1994}), 
or
as the \textit{geodesic equation}
\cite{Arnold1966};
$$
  \sdd{t}{} \dd{\xi^a}{l} = C_{da}^{b} \xi^d \dd{\xi^b}{l},
$$
where
the variables and coefficients
${\xi}^d$, $\dd{\xi^a}{l}$, $C_{da}^{b}$
respectively correspond to
$\GEV{V}{(\DCM{K}{})}$, $\GEV{M}{(\DCM{K}{})}$, 
and
$
  \tripleHMHD{}{\DCM{K}{}}{\DCM{P}{}}{\DCM{Q}{}}
  \EIGK(\DCM{K}{})/g(\DCM{K}{})
$
in the present study.
As is expected,
conservation laws for the energy and the mixed helicity are
easily proved from the symmetric properties of 
the obtained structure constant.

In the standard MHD limit $\HALL\to0$,
the eigenvalues, and thus,
the related quantities such as
the expansion coefficients, the Riemannian metric,
and
the structure constants,
diverge or shrink to zero 
unless $C_{i}=-C_{e}$.
Hence,
the DBF basis seems unsuitable for 
comparative analysis between HMHD and MHD in most cases.
However,
it is very interesting that
consideration of the effect of a uniform background magnetic field
yields 
such a linear wave equation that
uses the same DBF operator as
$
  C_{i}=-C_{e}=1
.
$
It is well-known that
the linear wave modes in the incompressible HMHD system
are
the ion cyclotron and whistler waves
and
are elegantly described by the GEV \cite{Galtier2006}.
That is,
the GEV is such that the DBF has non-diverging properties in the limit
$
  \HALL \to 0
.
$
Thus, 
among the wide variety of the DBF expansions,
the GEV expansion is the most suitable for
comparing the dynamics of the HMHD system to its MHD limit,
avoiding singularity problems.

Since the DBFs are constructed from 
the eigenfunctions of the curl operator,
it is easy to include Laplacian-type dissipation 
into the spectral representation of the equation of motion;
\begin{eqnarray}
&&
  \Big( g(\DCM{K}{}) \dd{t}{} + D(\DCM{K}{},\nu,\eta) \Big)
    \CC{\GEC{V}{(\DCM{K}{};t)}}
\nonumber\\&&\hspace{1em}
  =
  \sum_{\DCM{P}{}}\sum_{\DCM{Q}{}}
  \tripleHMHD{\big}{\DCM{K}{}}{\DCM{P}{}}{\DCM{Q}{}}\,
  \EIGK(\DCM{Q}{})\,
  \GEC{V}{(\DCM{P}{};t)}\,
  \GEC{V}{(\DCM{Q}{};t)}
,
\end{eqnarray}
where $D$, $\nu$, and $\eta$ are
the dissipation term coefficient given by
\begin{eqnarray}
  D(\DCM{K},\nu,\eta)
  :=
  {K^2}\left[
    \nu
    \left(\frac{1}{{\EIGK}(\DCM{K})}-\frac{\HEL_{K} K}{C_{e}}\right)^2
    +
    \frac{\eta}{C_{e}^2}
  \right]
,
\nonumber\\
\end{eqnarray}
the kinematic viscosity, and the resistivity, respectively.
This feature allows us
to apply the DBF expansion 
to such analyses as
the closure problem \cite{Galtier2006}
or
the direct numerical simulation \cite{ArakiMiura2015}.

\begin{acknowledgments}
The author expresses appreciation to the anonymous referees
for kind and fruitful comments 
and to Prof. H. Miura for his continuous encouragement.
This work was performed under the auspices of 
the NIFS Collaboration Research Program (NIFS13KNSS044, NIFS15KNSS065) 
and 
KAKENHI (Grant-in-Aid for Scientific Research(C)) 23540583.
The author would like to thank Enago for the English language review.
\end{acknowledgments}

\appendix
\section{Derivation of the Lin constraints\label{Derivation of Lin constraints}}

We briefly review the derivation of 
the formula for the variation of the tangent vector to an integral path.
For this derivation process,
we use only the exponential map and the Baker-Campbell-Hausdorff formula.
Since no material specific to any particular Lie algebra is used here,
the result is applicable to all the Lie groups.

\begin{figure}
\centering
\includegraphics[width=200pt]{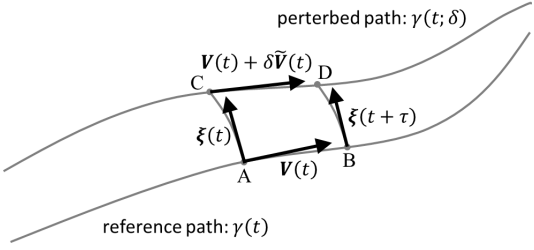}
\caption{\label{paths lin constraints}%
The derivation of the Lin constraints
using approximations of short intervals on paths.
}
\end{figure}

Let $\gamma(t;\delta)$ ($t\in[0,1]$) be
a path on $G$ with a variation parameter $\delta\in I\in\mathbb{R}$.
The path C$\to$D is approximated by
\begin{center}
$\gamma(t+\tau;\delta)$
$\approx$
$\exp\big[\tau\big(\bm{V}(t)+\delta\tilde{\bm{V}}(t)\big)\big]$
$\circ$
$\gamma(t;\delta)$,
\end{center}
where 
${\bm{V}}(t)$ 
is the tangent vector to the reference path ($\delta=0$), 
and
$\delta\tilde{\bm{V}}(t)$ is the small deviation.
The path
C$\to$A$\to$B$\to$D 
is%
, 
on the other hand,
also approximated by
\begin{eqnarray*}
  \gamma(t+\tau;\delta)
  & \approx &
  \exp\big(\delta\bm{\xi}(t+\tau)\big)
  \circ
  \exp\big(\tau\bm{V}(t)\big)
 \\&&
  \circ
  \exp\big(-\delta\bm{\xi}(t)\big)
  \circ
  \gamma(t;\delta)
\end{eqnarray*}
(see Figure \ref{paths lin constraints}).
Expanding
$
 \bm{\xi}(t+\tau)
 =
 \bm{\xi}(t)
 +\tau
 \dot{\bm{\xi}}(t)
 +o(\tau)
$
and
using the Baker-Campbell-Hausdorff formula at the lowest two orders,
we obtain
\begin{eqnarray}
&&
  \exp\big(\delta\bm{\xi}(t+\tau)\big)
  \circ
  \exp\big(\tau\bm{V}(t)\big)
  \circ
  \exp\big(-\delta\bm{\xi}(t)\big)
\nonumber\\
&&
  =
  \exp\Big[
    \tau\bm{V}(t)
    +
    \tau \delta \big(
      \dot{\bm{\xi}}(t)
      + 
      \LieBracket{}{\bm{\xi}(t)}{\bm{V}(t)}
    \big)
    +o(\delta)
    +o(\tau)
  \Big]
.
\nonumber\\
\end{eqnarray}
Since
the two approximated paths from C to D agree with each other 
in the limit $\delta\to0$ and $\tau\to0$,
we obtain the Lin constraints
\begin{eqnarray}
  \tilde{\bm{V}}(t)
  =
  \dot{\bm{\xi}}(t)
  + 
  \LieBracket{}{\bm{\xi}(t)}{\bm{V}(t)}
\label{lin constraints appendix}
\end{eqnarray}
at the order $O(\delta\tau)$.


\section{local expression of particle-relabeling symmetry
\label{sec:relabeling symmetry}}

By the term ``particle-relabeling symmetry,''
we recognize the invariance of the flow 
against the change of Lagrangian coordinates.
The freedom of choice of the action-preserving transformation exists 
only at the ``initial time'' 
and
the transformation along the integral path of the action is determined by 
this initial condition.
This symmetry is qualitatively different from the symmetry 
considered, for example, in gauge field theory, 
wherein, in principle, 
group transformation is applicable at any point in the relevant space and time
\cite{utiyama1956invariant}.

Thus,
the evolution of transformation should be considered.
Let $\bm{\xi}$ and $\epsilon$ be a displacement-generating vector field and 
a small parameter, respectively.
For an assigned flow and a sufficiently small displacement,
the displacement field must satisfy the particle tracing relation:
\begin{eqnarray}
  \vec{X}\big(\vec{a}+\epsilon\bm{\xi}(\vec{a};0);t\big)
  &=&
  \vec{X}(\vec{a};t)+\epsilon\bm{\xi}\big(\vec{X}(\vec{a};t);t\big)
,
\end{eqnarray}
where the $\vec{X}$ is the PTM for the assigned flow.
At the order $O(\epsilon)$, each component of $\bm{\xi}$ satisfies
\begin{eqnarray}
  {\xi}^{k}(\vec{a};0)\left.\dd{x^{k}}{X^{i}}\right|_{(\vec{a};t)}
  &=&
  {\xi}^{i}\big(\vec{X}(\vec{a};t);t\big)
.
\end{eqnarray}
Differentiating with respect to $t$ and evaluating at $t=0$, we obtain
\begin{eqnarray}
  {\xi}^{k}(\vec{a};0)\left.\dd{x^{k}\partial t}{^2X^{i}}\right|_{(\vec{a};0)}
  &=&
  \left.\dd{x^{k}}{{\xi}^{i}}\right|_{(\vec{X}(\vec{a};0);0)}
  \left.\dd{t}{X^{k}}\right|_{(\vec{a};0)}
 \nonumber\\&&
  +
  \dd{t}{{\xi}^{i}}_{(\vec{X}(\vec{a};0);0)}
.
\end{eqnarray}
The relation (\ref{Eulerian Lagrangian velocity}) leads to
the following PDE for the vector fields in the Eulerian specification:
\begin{eqnarray}
  \left(
    \dd{t}{{\xi}^{i}}
    +
    V^{k} \dd{x^{k}}{{\xi}^{i}}
    -
    {\xi}^{k} \dd{x^{k}}{V^{i}}
  \right)_{(\vec{X}(\vec{a};0);0)}
  =
  0
,
\end{eqnarray}
which is the evolution equation for the frozen-in line element.
Since the Lie bracket of the vector fields is given by 
(\ref{Lie bracket and Lie derivative}),
the obtained evolution equation agrees with (\ref{lin constraints appendix})
for $\tilde{\bm{V}}(t)=\bm{0}$.
For divergence-free fields in a three-dimensional space,  
the equation is rewritten using vector analysis notation as
\begin{eqnarray}
  \partial_t\bm{\xi}+\curl(\bm{\xi}\times\bm{V})=\bm{0}
.
\end{eqnarray}

\section{Structure constants for the DBF basis
\label{derivation str const}}
\renewcommand{\DCM}[1]{\tilde{#1}}
The symbol $\tripleHMHD{\big}{\DCM{K}{}}{\DCM{P}{}}{\DCM{Q}{}}$,
which is related to the structure constant of the Lie algebra,
is defined by using a combination of the Riemannian metric and 
the Poisson bracket for the HMHD system as follows:
\begin{eqnarray}
&&\hspace{-2em}
  \tripleHMHD{\big}{\DCM{K}{}}{\DCM{P}{}}{\DCM{Q}{}}\,
  {\EIGK}(\DCM{K}{})
  :=
  \BraKet{\Big}{
    \GEV{\Psi}{}{(\DCM{K}{})}
  }{
    \LieBracket{\big}{
      \GEV{\Psi}{}{(\DCM{P}{})}
    }{
      \GEV{\Psi}{}{(\DCM{Q}{})}
    }
  }
\nonumber\\
  &=&
  \ParKet{\Big}{
    \U{\bm{\Psi}_{i}}(\DCM{K}{})
  }{
    \curl\big(
      \bm{\Psi}_{i}(\DCM{P}{})\times\bm{\Psi}_{i}(\DCM{Q}{})
    \big)
  }
 \nonumber\\&&
  +
  \ParKet{\Big}{
    \U{\bm{\Psi}_{e}}(\DCM{K}{})
  }{
    \curl\big(
      \bm{\Psi}_{e}(\DCM{P}{})\times\bm{\Psi}_{e}(\DCM{Q}{})
    \big)
  }
\nonumber\\
  &=&
  \frac{{\EIGK}(\DCM{K}{})}{\HALL}
  \int {\rm{d}}^3\vec{x}\Big[
    C_{i}^{-1}
    \bm{\Psi}_{i}(\DCM{K}{})\cdot\big(
      \bm{\Psi}_{i}(\DCM{P}{})\times\bm{\Psi}_{i}(\DCM{Q}{})
    \big)
   \nonumber\\&&\hspace{5em}
    +
    C_{e}^{-1}
    \bm{\Psi}_{e}(\DCM{K}{})\cdot\big(
      \bm{\Psi}_{e}(\DCM{P}{})\times\bm{\Psi}_{e}(\DCM{Q}{})
    \big)
  \Big]
,
\hspace{1em}
\end{eqnarray}
where the second line is derived 
by integrating by parts with respect to $\vec{x}$
and
by using the relation (\ref{Psi and underline Psi}).
Substitution of (\ref{base functions V}) into the second line
yields (\ref{str. const. core DCB}),
i.e.,
the explicit expression of the symbol
$
  \tripleHMHD{\big}{\DCM{K}{}}{\DCM{P}{}}{\DCM{Q}{}}
.
$
\renewcommand{\sakujo}[1]{}\sakujo{
Thus
we obtain the value of the structure constant of the Lie algebra
with respect to the double Beltrami base functions as
\begin{eqnarray}
&&
  \frac{
    \BraKet{\Big}{\GEV{\Psi}{}{(\DCM{K}{})}}{
      \LieBracket{\big}{
        \GEV{\Psi}{}{(\DCM{P}{})}
      }{
        \GEV{\Psi}{}{(\DCM{Q}{})}
      }
    }
  }{
    \BraKet{\Big}{\GEV{\Psi}{}{(\DCM{K}{})}}{\GEV{\Psi}{}{(\DCM{K}{})}}
  }
  =
  \frac{
    \tripleHMHD{\big}{\DCM{K}{}}{\DCM{P}{}}{\DCM{Q}{}}
    {\EIGK}(\DCM{K}{})
  }{
    g(\DCM{K}{})
  }
.
\label{str const in Lambda}
\end{eqnarray}
}
%
%
Using this symbol, the first variation of action reads as
\begin{eqnarray}
&&
  \int_0^1dt
  \BraKet{\Big}{
    \sum_{\DCM{Q}}\GEV{V}{}(\DCM{Q}{})
  }{
    \partial_{t}\GEV{\xi}{}(\DCM{K}{})
    +
    \LieBracket{\big}{
      \GEV{\xi}{}(\DCM{K}{})
    }{
      \sum_{\DCM{P}}\GEV{V}{}{(\DCM{P}{})}
    }
  }
\nonumber\\&&
  =
  \int_0^1dt\Big(
  g(\DCM{K})
  \CC{\GEC{V}{}(\DCM{K}{})}
  \partial_t{\GEC{\xi}{}}(\DCM{K}{})
 \nonumber\\&&\hspace{4em}
  +
  \sum_{\DCM{P},\DCM{Q}}
  \EIGK(\DCM{Q}{})
  \tripleHMHD{\big}{\DCM{Q}{}}{\DCM{K}{}}{\DCM{P}{}}
  \GEC{V}{}(\DCM{Q}{})
  {\GEC{\xi}{}}(\DCM{K}{})
  \GEC{V}{}(\DCM{P}{})
  \Big)
,
\nonumber\\&&
  =
  \left(
  g(\DCM{K})
  \CC{\GEC{V}{}(\DCM{K}{})}
  {\GEC{\xi}{}}(\DCM{K}{})
  \right)_{t=1}
  -
  \left(
  g(\DCM{K})
  \CC{\GEC{V}{}(\DCM{K}{})}
  {\GEC{\xi}{}}(\DCM{K}{})
  \right)_{t=0}
 \nonumber\\&&\hspace{1em}
  +
  \int_0^1dt{\GEC{\xi}{}}(\DCM{K}{})\Big(
  -
  g(\DCM{K})
  \partial_t\CC{{\GEC{V}{}}(\DCM{K}{})}
 \nonumber\\&&\hspace{5em}
  +
  \sum_{\DCM{P},\DCM{Q}}
  \EIGK(\DCM{Q}{})
  \tripleHMHD{\big}{\DCM{K}{}}{\DCM{P}{}}{\DCM{Q}{}}
  \GEC{V}{}(\DCM{Q}{})
  \GEC{V}{}(\DCM{P}{})
  \Big)
.
\hspace{1em}
\end{eqnarray}
We obtain the Euler-Lagrange equation 
(\ref{eq of mot in double Beltram coeff})
if the fixed path end conditions are imposed.
\bibliography{apssamp}

\end{document}